\documentclass[preprint2]{aastex}

\shorttitle{Evolution of the \ha{} LF}
\shortauthors{Westra et al.}

\providecommand{\includeIDLfigP}[2][\columnwidth]{\includegraphics[angle=90, width=#1, trim=6pt 16pt 18pt 15pt, clip]{#2}}
\providecommand{\includeIDLfigPcustom}[6][\columnwidth]{\includegraphics[angle=90, width=#1, trim=#2 #3 #4 #5, clip]{#6}}
\newcommand{\lya}{Ly$\alpha$}
\newcommand{\ha}{\ifmmode{\mathrm{H\alpha}}\else{H$\alpha$}\fi}
\newcommand{\hb}{\ifmmode{\mathrm{H\beta}}\else{H$\beta$}\fi}
\newcommand{\hc}{\ifmmode{\mathrm{H\gamma}}\else{H$\gamma$}\fi}

\newcommand{\oii}{[{\sc Oii}]}
\newcommand{\oiii}{\ifmmode{\mbox{[{\sc O\,iii}]}}\else{[{\sc O\,iii}]}\fi}

\newcommand{\Oiiib}{\oiii~$\lambda$5007}
\newcommand{\nii}{\ifmmode{\mbox{[{\sc N\,ii}]}}\else{[{\sc N\,ii}]}\fi}
\newcommand{\Nii}{\nii~$\lambda\lambda$6550,6585}

\newcommand{\Niib}{\nii~$\lambda$6585}
\newcommand{\sii}{[{\sc S\,ii}]}
\newcommand{\Sii}{\sii~$\lambda\lambda$6733,6718}

\newcommand{\kms}{\ifmmode{\mathrm{km\,s^{-1}}}\else{km\,s$^{-1}$}\fi}
\newcommand{\perMpcSq}{\ifmmode{\mathrm{Mpc^{-2}}}\else{Mpc$^{-2}$}\fi}
\newcommand{\perMpcQube}{\ifmmode{\mathrm{Mpc^{-3}}}\else{Mpc$^{-3}$}\fi}
\newcommand{\perMpcQ}{\perMpcQube}
\newcommand{\MpcQ}{\ifmmode{\mathrm{Mpc^{3}}}\else{Mpc$^{3}$}\fi}
\newcommand{\uJy}{\ifmmode{\mathrm{\mu Jy}}\else{$\mu$Jy}\fi}
\newcommand{\kmsMpc}{\ifmmode{\mathrm{km\,s^{-1}\,Mpc^{-1}}}\else{km\,s$^{-1}$\,Mpc$^{-1}$}\fi}
\newcommand{\ergs}{\ifmmode{\mathrm{erg\,s^{-1}}}\else{erg\,s$^{-1}$}\fi}

\newcommand{\fluxunits}{\ergs\,\ifmmode{\mathrm{cm^{-2}}}\else{cm$^{-2}$}\fi}
\newcommand{\ergsPerAng}{\fluxunits\,\ifmmode{\mathrm{\AA^{-1}}}\else{\AA$^{-1}$}\fi}
\newcommand{\ergsPerHz}{\fluxunits\,\ifmmode{\mathrm{Hz^{-1}}}\else{Hz$^{-1}$}\fi}
\newcommand{\Msunyr}{\ifmmode{\mathrm{M_\odot\,yr^{-1}}}\else{M$_\odot$\,yr$^{-1}$}\fi}
\newcommand{\lumDens}{\ergs\,\ifmmode{\mathrm{Mpc^{-3}}}\else{Mpc$^{-3}$}\fi}
\newcommand{\sfDens}{\Msunyr\,\ifmmode{\mathrm{Mpc^{-3}}}\else{Mpc$^{-3}$}\fi}

\providecommand{\pow}[2][10]{#1^{#2}}
\providecommand{\eqref}[1]{(\ref{#1})}

\providecommand{\giCol}{\ifmmode(g'-i')\else$(g'-i')$\fi}
\providecommand{\VRCol}{\ifmmode(V-R)\else$(V-R)$\fi}

\begin{document}

\defcitealias{Shioya08}{S08}

\title{Evolution of the \ha{} luminosity function}

\author{Eduard Westra\altaffilmark{1},
  Margaret~J.~Geller\altaffilmark{1},
  Michael~J.~Kurtz\altaffilmark{1},
  Daniel~G.~Fabricant\altaffilmark{1},
  Ian~Dell'Antonio\altaffilmark{2}}

\altaffiltext{1}{Smithsonian Astrophysical Observatory, 60 Garden
Street, Cambridge, MA 02138, USA}

\altaffiltext{2}{Brown University, Department of Physics, Box 1843,
Providence, RI 02912, USA}

\email{ewestra@cfa.harvard.edu}

\begin{abstract}
The Smithsonian Hectospec Lensing Survey (SHELS) is a window on the
star formation history over the last 4\,Gyr. SHELS is a
spectroscopically complete survey for $R_\mathrm{tot} < 20.3$ over
4\,\sq{}\degr{}. We use the 10k spectra to select a sample of pure
star forming galaxies based on their \ha{} emission line. We use the
spectroscopy to determine extinction corrections for individual
galaxies and to remove active galaxies in order to reduce systematic
uncertainties. We use the large volume of SHELS with the depth of a
narrowband survey for \ha{} galaxies at $z \sim 0.24$ to make a
combined determination of the \ha{} luminosity function at $z \sim
0.24$. The large area covered by SHELS yields a survey volume big
enough to determine the bright end of the \ha{} luminosity function
from redshift 0.100 to 0.377 for an assumed fixed faint-end slope
$\alpha = -1.20$. The bright end evolves: the characteristic
luminosity $L^*$ increases by 0.84\,dex over this redshift
range. Similarly, the star formation density increases by
0.11\,dex. The fraction of galaxies with a close neighbor increases by
a factor of $2-5$ for $L_\ha \gtrsim L^*$ in each of the
redshift bins. We conclude that triggered star formation is an
important influence for star forming galaxies with \ha{} emission.
\end{abstract}

\keywords{galaxies: evolution -- galaxies: interactions -- galaxies:
  luminosity function -- galaxies: starbursts}

\section{Introduction}
\label{sec:intro}
Determining the star formation history of the Universe is a crucial
part of understanding the formation and evolution of
galaxies. Exploration of the global star formation history has two
components: ({\it i}) measurement of the star formation density over
time and ({\it ii}) understanding the physical processes that drive
star formation. Here we use a large, moderate-depth spectroscopic
survey to address both issues: ({\it i}) we determine the star
formation density over the last 4\,Gyr using the \ha{} emission line
as star formation indicator and ({\it ii}) we investigate the
possible influence of galaxy interactions on the \ha{} luminosity
function.

There is abundant observational evidence for an order of magnitude
increase in the star formation density since redshift $z\sim 1-2$
\citetext{\citealp{Lilly96,Madau96}; and the compilations of
\citealp{Hopkins04,Hopkins06}}. Major mergers, tidal interactions, gas
removal from conversion into stars, and/or ram pressure stripping may
explain the decrease in the star formation. The challenge is deciding
which of these processes are important in quenching of star formation
\citep{Bell05}.

The decline in star formation density coincides with a rapid decrease
in the characteristic luminosity of galaxies ($L^*$) in the rest-frame
$U$-band \citep[e.g.][]{Ilbert05,Prescott09}. A decrease in the number
of merging systems can explain the decrease of the characteristic
luminosity $L^*$ \citep{Fevre00}. \citet{Sobral09} find a strong
morphology-\ha{} luminosity relation for mergers and non-mergers. The
characteristic luminosity $L^*$ defines a critical switch-over
luminosity between the mergers and non-mergers; the mergers are more
luminous.

Studies of close pairs show that enhancement in the star formation
rate is largest for galaxies in major pairs \citep[$|\Delta m|
\lesssim 0.7 - 2$;][]{Woods06,Woods07,Ellison08} and that the average
star formation rate in a galaxy increases with decreasing projected
separation \citep{Li08}. Simulations of interacting and merging
galaxies reveal that the interactions can trigger short powerful
bursts of star formation by forcing substantial fractions of the gas
into the central regions \citep{Mihos96}.

Systematic effects dominate the comparison of star formation rates
determined from different star formation indicators like the
rest-frame ultra-violet (UV) and \ha{}. Hence, to study the variation
of the star formation density with time, the use of a single star
formation indicator is best. The rest-frame UV spectrum of a galaxy
directly measures the population of newborn stars
\citep[e.g.][]{Lilly96,Treyer98}. However, the rest-frame UV is
strongly attenuated \citep[e.g.][]{Cardelli89,Calzetti00}. The
most-direct optical indicator is the \ha{} emission line emitted by
gas surrounding the embedded star forming region
\citep[e.g.][]{Kennicutt98}. The \ha{} line is also affected by
attenuation--albeit less than the UV--which can be corrected using
spectroscopy.

Many surveys use narrowband filters
\citep{Thompson96,Moorwood00,Jones01,Fujita03,Hippelein03,Ly07,Pascual07,Dale08,Geach08,Morioka08,Shioya08,Westra08,Sobral09}
to determine the \ha{} luminosity function parameters over a range of
redshifts. Despite the depth of the narrowband surveys, measurements
of individual luminosity function parameters and the star formation
density are not well-constrained.

Narrowband surveys lack spectroscopy for the faint \ha{} emitting
galaxies. Thus, general assumptions about stellar absorption,
extinction corrections, contributions by active galactic nuclei
(AGNs), or interloper contamination need to be made for the sample as
a whole rather than for each galaxy. These issues may lead to
systematic uncertainties. \citet{Massarotti01} show that applying an
average extinction correction introduces a systematic underestimate of
the extinction-corrected star formation density. A spectroscopic
survey does not suffer these limitations, although it is usually
limited in its depth.

Several spectroscopic \ha{} surveys exist
\citep[e.g.][]{Gallego95,Tresse98,Sullivan00,Tresse02,Perez03,Shim09}.
Both \citeauthor{Gallego95} and \citeauthor{Perez03} use the
Universidad Complutense de Madrid (UCM) survey. This survey covers an
extremely wide area on the sky (472\,\sq{}\degr{}). However, it is
limited to a very low redshift ($z_\mathrm{max} \sim 0.045$). For
their \ha{} survey, \citet{Sullivan00} use galaxies selected from UV
imaging in a 2.2\,\sq{}\degr{} field. The other surveys have an area
$\le$ 0.25\,\sq{}\degr{}. Thus, most surveys are too limited in volume
to overcome cosmic variance.

The Smithsonian Hectospec Lensing Survey (SHELS) is a spectroscopic
survey covering 4\,\sq{}\degr{} on the sky to a limiting $R$-band
magnitude $R_\mathrm{tot} = 20.3$ \citep{Geller05}. We use SHELS to
obtain a consistent determination of the star formation history over
the last 4\,Gyr based on the \ha{} emission line over a relatively
large area and redshift range.

The spectroscopy enables us to reduce systematic uncertainties by
allowing an individual galaxy extinction correction. We can also
remove individual AGNs rather than applying a global correction factor
for contamination by AGNs as is done in narrowband surveys. We use the
large survey area to determine the characteristic luminosity $L^*$ of
the \ha{} luminosity function and associated systematic uncertainties.

We discuss the SHELS spectroscopic data in
Section~\ref{sec:fieldselection}. In Section~\ref{sec:sampleselection}
we introduce our \ha{} sample selection. We combine our $R$-band
selected \ha{} sample with the narrowband \ha{} survey of
\citet{Shioya08} in Section~\ref{sec:nbvbb} to obtain a
jointly-determined \ha{} luminosity function at
$z\sim0.24$. Sections~\ref{sec:halfSHELS} and \ref{sec:sfd} discuss
the derivation and evolution of the luminosity function and star
formation density, respectively, over the past 4\,Gyr. We include an
investigation of the influence of our selection criteria on the
derivation of the luminosity function parameters. In
Section~\ref{sec:properties} we examine the stellar age of the star
forming galaxies and the influence of galaxy-galaxy interactions on
these galaxies. We summarize our results in Section~\ref{sec:summary}.

Throughout this paper we assume a flat Universe with $H_0 =
71$\,\kmsMpc{}, $\Omega_{\rm m} = 0.27$ and $\Omega_{\Lambda} =
0.73$. All quoted magnitudes are on the AB-system and luminosities are
in \ergs{}.

\section{SHELS observations}
\label{sec:fieldselection}
We constructed the SHELS galaxy catalog from the $R$-band source list
for the F2 field of the Deep Lens Survey \citep{Wittman02,Wittman06}.
The DLS is an NOAO key program covering 20\,\sq{}\degr{} in five
separate fields; the 4.2\,\sq{}\degr{} F2 field is centered at $\alpha
= 09^h19^m32.4^s$ and $\delta = +30\degr{}00\arcmin{}00\arcsec{}$. We
exclude regions around bright stars ($\sim$\,5\,\% of the total
survey) resulting in an effective area of 4.0\,\sq{}\degr{}. We use
surface brightness and magnitude to separate stars from galaxies. This
selection removes some AGN.

Photometric observations of F2 were made with the MOSAIC I imager
\citep{Muller98} on the KPNO Mayall 4\,m telescope between 1999
November and 2004 November. The $R$-band exposures, all taken in
seeing $ < 0.9\arcsec{}$ FWHM, are the basis for the SHELS survey. The
effective exposure time is about 14,500 seconds and the 1\,$\sigma$
surface brightness limit in $R$ is 28.7 magnitudes per square
arcsecond. \citet{Wittman06} describe the reduction pipeline.

We acquired spectra for the galaxies with the Hectospec fiber-fed
spectrograph \citep{Fabricant98,Fabricant05} on the MMT from 2004
April 13 to 2007 April 20. The spectrograph is fed by 300 fibers that
can be positioned over a 1\degr{} field. Roughly 30 fibers per
exposure are used to determine the sky. The Hectospec observation
planning software \citep{Roll98} enables efficient acquisition of a
magnitude limited sample.

The SHELS spectra cover the wavelength range $\lambda = 3,500 -
10,000$\,\AA{} with a resolution of $\sim$6\,\AA{}. Exposure times
ranged from 0.75~hours to 2~hours for the lowest surface brightness
objects in the survey. We reduced the data with the standard Hectospec
pipeline \citep{Mink07} and derived redshifts with RVSAO
\citep{Kurtz98} with templates constructed for this purpose
\citep{Fabricant05}. We have 1,468 objects that have been observed
twice. These repeat observations imply a mean internal error of
56\,\kms{} for absorption-line objects and 21\,\kms{} for
emission-line objects \citep[see also][]{Fabricant05}.

\citet{Fabricant08} describe the technique we use for photometric
calibration of the Hectospec spectra based on the particularly stable
instrument response. For galaxies in common between SHELS and SDSS,
the normalized \ha{} line fluxes agree well in spite of the difference
in fiber diameters for the Hectospec (1\farcs5) and the SDSS
(3\arcsec{}). For high-signal-to-noise SHELS spectra, the typical
uncertainties in emission line fluxes are 18\,\%.

SHELS includes 9,825 galaxies to the limiting apparent magnitude. The
overall completeness of the redshift survey to a total\footnote{The
total magnitude is the SExtractor \citep{Bertin96} {\sc mag\_auto} as
opposed to an aperture magnitude.} $R$-band magnitude of
$R_\mathrm{tot} \le 20.3$ is 97.7\,\%, i.e. 9,595 galaxies have a
redshift measured; the differential completeness at the limiting
magnitude is 94.6\,\%. The 230 objects without redshifts are low
surface and/or faint objects, or objects near the survey corners and
edges. M.~J.~Kurtz et al. (2010; in preparation) includes a detailed
description of the full redshift survey.

The SHELS survey also includes 1,852 galaxies with $20.3 < R \leq
20.6$, for which we have measured a redshift; the total sample of
galaxies with $20.3 < R \leq 20.6$ is 3,590, i.e. the survey is 52\,\%
complete in this magnitude interval. The completeness is patchy across
the field.

\begin{figure*}[tb]
  \centering
  \includeIDLfigPcustom[\textwidth]{12pt}{20pt}{112pt}{8pt}{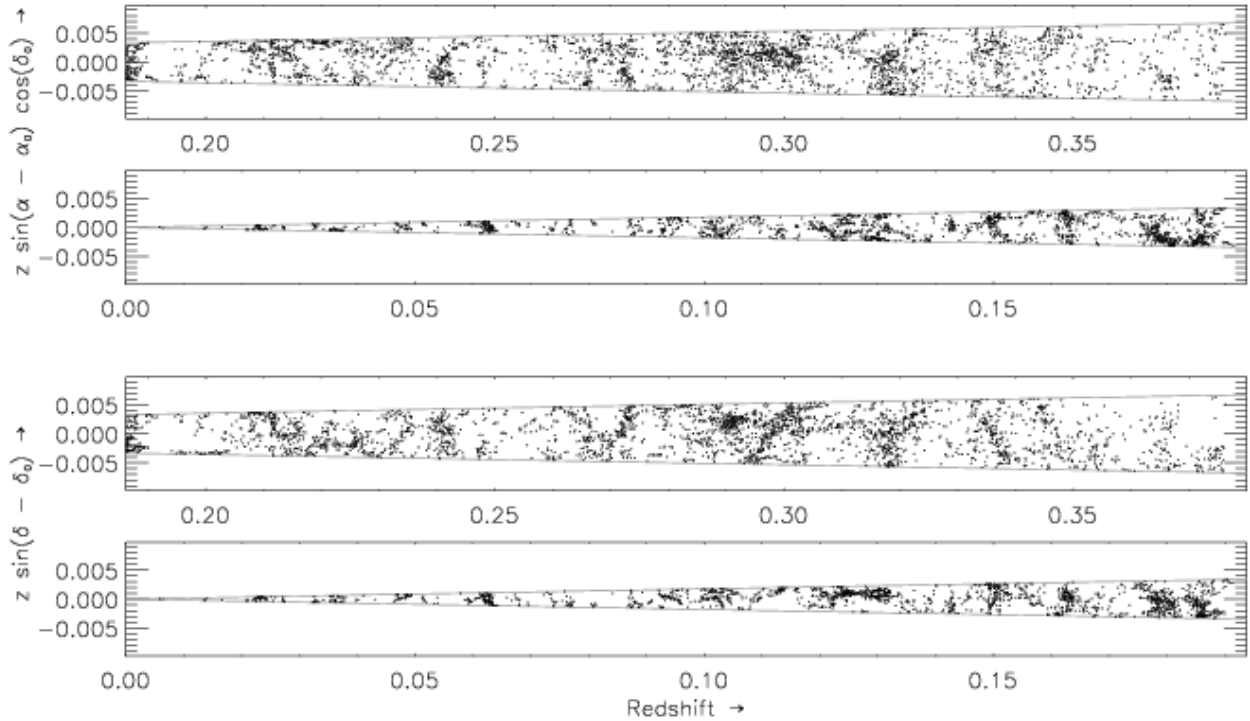}
  \caption{Redshift cone diagram for the galaxies in the final sample:
  $R_\mathrm{tot} \leq 20.3$, S/N$_\mathrm{\ha{}} > 5$ and
  $f_\mathrm{\ha} \ge \pow{-15.5}$\,\fluxunits{}. AGNs have not been
  removed from this sample. The large-scale structure is apparent with
  extended low-density regions and well-populated narrow structures.}
  \label{fig:cone}
\end{figure*}

The F2 field contains an atypical under-dense region at the lowest
redshifts because the DLS fields are selected against nearby clusters
at $z < 0.1$. We show the redshift distribution of our \ha{} galaxies
in Figure~\ref{fig:cone}.

\subsection{The $R$-band $k+e$-corrections}
To calculate the absolute $R$-band magnitude $M_R$ we determine the
appropriate $k+e$-corrections. The $k+e$-correction converts the
observed absolute magnitude to the rest-frame of the galaxy,
correcting for redshift and evolution. We use the $k+e$-corrections
for 9 types of galaxies: bright cluster (BCG), elliptical (E), S0, Sa,
Sb, Sbc, Sc, Sd and irregular (Irr) galaxies determined by J.~Annis
\citetext{priv.~comm.}\footnote{The table with the corrections for the
SDSS filter set as function of galaxy type and redshift can be
obtained from
\url{http://home.fnal.gov/~annis/astrophys/kcorr/kcorr.html}.}. We
use the corrections for the SDSS $r'$-filter as a function of redshift
and \giCol{}-color because the SDSS $r'$-filter is similar to the
$R$-filter used for the DLS. We obtain \giCol{} by cross-matching our
catalog with SDSS DR6 \citep{SDSS6}. For those galaxies not found in
SDSS DR6 (these galaxies are either unresolved or below the surface
brightness limit in SDSS) we convert the \VRCol{} from 61 galaxies in
the DLS to \giCol{}. For 42 galaxies we cannot determine or derive
\giCol{} due to the proximity of another object; we assume that these
are Sa galaxies. We interpolate the models in redshift to obtain the
$k+e$-corrections for each galaxy type determined by its \giCol{} and
redshift.

\section{\ha{} sample selection}
\label{sec:sampleselection}
We use SHELS to construct \ha{} luminosity functions over the redshift
range $0.010 < z < 0.377$ (Table~\ref{tab:ngals}). Here, we describe
the determination of our final emission-line luminosities and the
discrimination between pure star-forming galaxies and AGNs.

\subsection{Emission-line measurements}
\label{sec:contsub}
\label{sec:emabmeasure}
The emission-line flux emanating from star-forming regions is affected
by the absorption-line spectrum from the underlying stellar
population. The absorption mostly affects the measurements of the
hydrogen Balmer lines. To measure the emission-line flux we thus
remove the contribution of the stellar population.

We use the \citet{Tremonti04} continuum subtraction method to correct
for the stellar absorption rather than applying a constant, global
correction \citep[e.g.,][]{Hopkins03}. The \citeauthor{Tremonti04}
method removes the stellar continuum by fitting a linear combination
of template spectra resampled to the correct velocity dispersion. The
method also accounts for redshift and reddening. The template spectra
are based on single stellar population models generated by the
population synthesis code of \citet{Bruzual03}. We use models with 10
different ages (0.005, 0.025, 0.1, 0.3, 0.6, 0.9, 1.4, 2.5, 5 and 10
Gyr) at solar metallicity.

We determine the emission-line fluxes from the continuum-subtracted
spectra by integrating the line flux within a top-hat filter centered
on the emission-line. We remove any local over- or under-subtraction
of the continuum by subtracting the mean of the flux-density at both
sides of the filter. Next, we determine the continuum level by taking
the mean of the flux-density at wavelengths bluer and redder than the
emission-line on the best-fit continuum model. Finally, we determine
the absorption contribution of the underlying stellar population using
the same top-hat filter but on the best-fit model; we remove the flux
contributed by the continuum.

\begin{figure}[tb]
  \centering
  \includeIDLfigPcustom{15pt}{15pt}{26pt}{7pt}{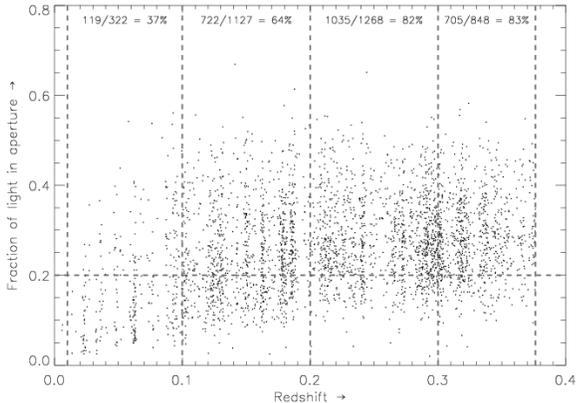}
  \caption{Fraction of light contained in the 1\farcs5 fibers as a
  function of redshift. We indicate the fraction of galaxies with more
  than 20\,\% of the light contained in the fiber ({\it horizontal
  dashed line}) for each redshift bin ({\it vertical dashed lines}) we
  use to construct the \ha{} luminosity functions. The galaxies have
  $R_\mathrm{tot} \leq 20.3$, S/N$_\mathrm{\ha{}} > 5$ and
  $f_\mathrm{\ha} \ge \pow{-15.5}$\,\fluxunits{}. The AGNs have been
  removed.}
  \label{fig:lightfraction}
\end{figure}

The Hectospec fibers have a fixed diameter of 1\farcs5. At all
redshifts where \ha{} is observable (and in particular at the lowest
redshifts) the fiber does not cover the entire galaxy. Hence, we use
an aperture correction
\label{sec:apcorr}
\begin{equation}
  A = \pow{-0.4(m_\mathrm{total} - m_\mathrm{fiber})}
  \label{eq:apcorr}
\end{equation}
to correct for the fiber-covering
fraction. Figure~\ref{fig:lightfraction} shows the fraction of light,
$1/A$, contained in the fiber as a function of redshift.

\citet{Kewley05} show that a spectrum measuring at least 20\,\% of the
galaxy light avoids substantial scatter between the nuclear and
integrated SFR measurements. The overall majority of galaxies from
SHELS have a light-fraction $1/A \ge 20\,\%$
(Figure~\ref{fig:lightfraction}).

\citet{Fabricant08} compared the \ha{} and \oii{} emission-line fluxes
from SHELS with SDSS DR6 after making an aperture correction. They
found excellent agreement between the two surveys, even though the
fibers of the SDSS spectrograph are 3\arcsec{} in diameter. Moreover,
most of the SDSS galaxies are at low redshift ($z \lesssim 0.14$)
where we have the largest fraction of galaxies with a light-fraction
less than 20\,\%.

There is no dependence of final \ha{} luminosity on the
light-fraction. We are thus confident that the use of these aperture
corrections does not affect the final results even when the covering
fraction is small.

\begin{figure}[tb]
  \centering
  \includeIDLfigP{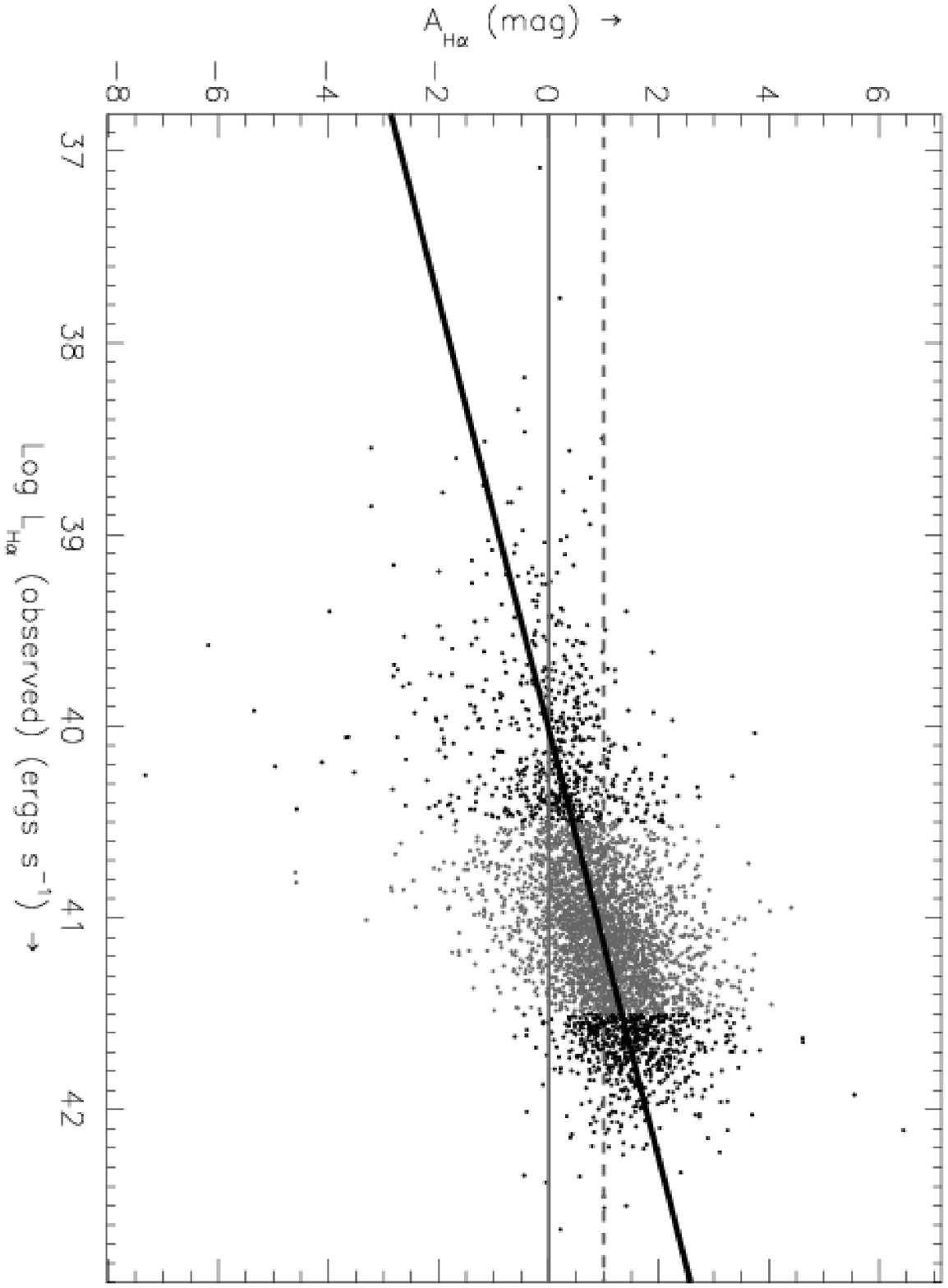}
  \caption{Attenuation at \ha{} as function of observed \ha{}
  luminosity. The black line indicates the least-absolute-deviates fit
  to the gray points. We indicate $A_\ha{} = 0$ ({\it solid horizontal
  line}) and a commonly assumed value of $A_\mathrm{\ha}$
  \citep[$A_\mathrm{\ha} = 1$, {\it dashed horizontal line};
  e.g.][]{Tresse98,Fujita03,Ly07,Sobral09}}
  \label{fig:extinction}
\end{figure}

\subsection{Extinction correction}
\label{sec:extcorr}
The light from star forming regions in a galaxy is often heavily
attenuated. To determine the intrinsic SFR of a galaxy we must remove
the effects of attenuation.

We calculate the attenuation by comparing the observed value of the
Balmer decrement (corrected for stellar absorption) with the
theoretical value \citep[$f_\ha/f_\hb = 2.87$ for $T\,=\,\pow{4}$\,K
and case B recombination; Table~2 of][]{Calzetti01}. The intrinsic
flux is $f_\mathrm{intr}(\lambda) = f_\mathrm{obs}(\lambda) \pow{0.4
A_\lambda}$, where $A_\lambda$ is the wavelength-dependent
extinction. $A_\lambda$ is
\begin{eqnarray}
  A_\lambda & = & k(\lambda) E(B-V)_\mathrm{gas} \nonumber\\
  & = & k(\lambda)\frac{2.5 \log R_{\alpha \beta}}{k(\hb) - k(\ha)}~,
\end{eqnarray}
where $R_{\alpha\beta}$ is the ratio of the attenuated-to-intrinsic
Balmer line ratios, $k(\hb) - k(\ha)$ is the differential extinction
between the wavelengths of \hb{} and \ha{}, and $k(\lambda)$ is the
extinction at wavelength $\lambda$. We apply the \citet{Calzetti00}
extinction law, which has $k(V) = 4.05$, $k(\ha) = 3.325$ and $k(\hb)
= 4.596$.

\begin{figure}[tb]
  \centering
  \includeIDLfigPcustom{16pt}{16pt}{30pt}{21pt}{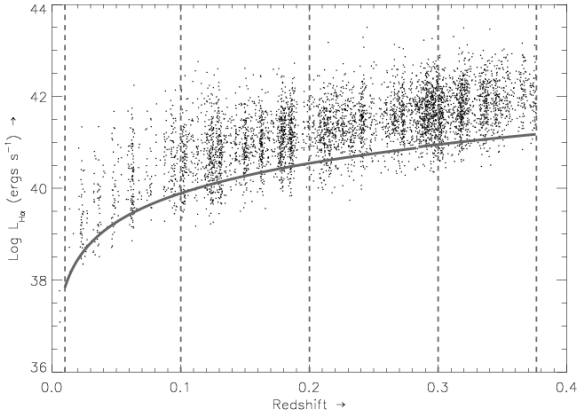}
  \caption{\ha{} luminosity (corrected for extinction) as a function
  of redshift. The solid line indicates the additional selection
  criterion $f_\mathrm{\ha} \ge \pow{-15.5}$\,\fluxunits{}. Vertical
  dashed lines show the edges of the redshift bins used to construct
  the SHELS \ha{} luminosity functions.}
  \label{fig:lhavz}
\end{figure}

Figure~\ref{fig:extinction} shows the attenuation as a function of
observed \ha{} luminosity for galaxies with both \ha{} and \hb{} at a
$\mathrm{S/N} > 5$. We use the relation between $A_\ha{}$ and
$L_\ha{}$ as determined from a least-absolute-deviates fit to the
high-S/N data-points with observed luminosities $40.5 < \log
L_\mathrm{\ha} < 41.5$ ({\it gray points}) for galaxies where
$\mathrm{S/N}_\mathrm{\hb} \le$ 5 or where the observed equivalent
width of \hb{}, OEW$_\mathrm{\hb} \le$ 1\,\AA{} (uncorrected for
stellar absorption). We limit OEW$_\mathrm{\hb}$ to avoid galaxies
with excessively large attenuation resulting from a very small
(noise-dominated) \hb{} flux compared to \ha{}. We assume that
galaxies with $A_\ha{} \le 0$ to have no attenuation and assign
$A_\ha{} = 0$ to these galaxies.

\subsection{Sample definition}
\label{sec:sampledef}
Figure~\ref{fig:lhavz} shows the \ha{} luminosity as a function of
redshift. Below $f_\mathrm{\ha} = \pow{-15.5}$\,\fluxunits{} the
number of galaxies decreases rapidly. We impose a constant \ha{} flux
limit $f_\mathrm{\ha} \ge \pow{-15.5}$\,\fluxunits{} on the sample
(after corrections for stellar absorption and attenuation) because we
are only complete to this flux. We apply this criterion in addition to
the magnitude limit ($R_\mathrm{tot} \leq 20.3$) and the
$\mathrm{S/N}_\mathrm{\ha} > 5$ requirement.

\subsection{AGN classification}
\label{sec:agn}
The presence of an active nucleus in a galaxy contributes to the
(apparent) star-formation in the galaxy. For example,
\citet{Pascual01} find that approximately 15\,\% of the luminosity
density of the UCM survey \citep{Gallego95} results from galaxies
identified as AGN. \citet{Westra08} find a 5\,\% contribution for
their survey.

\begin{figure}[tb]
  \centering
  \includeIDLfigPcustom{6pt}{16pt}{18pt}{5pt}{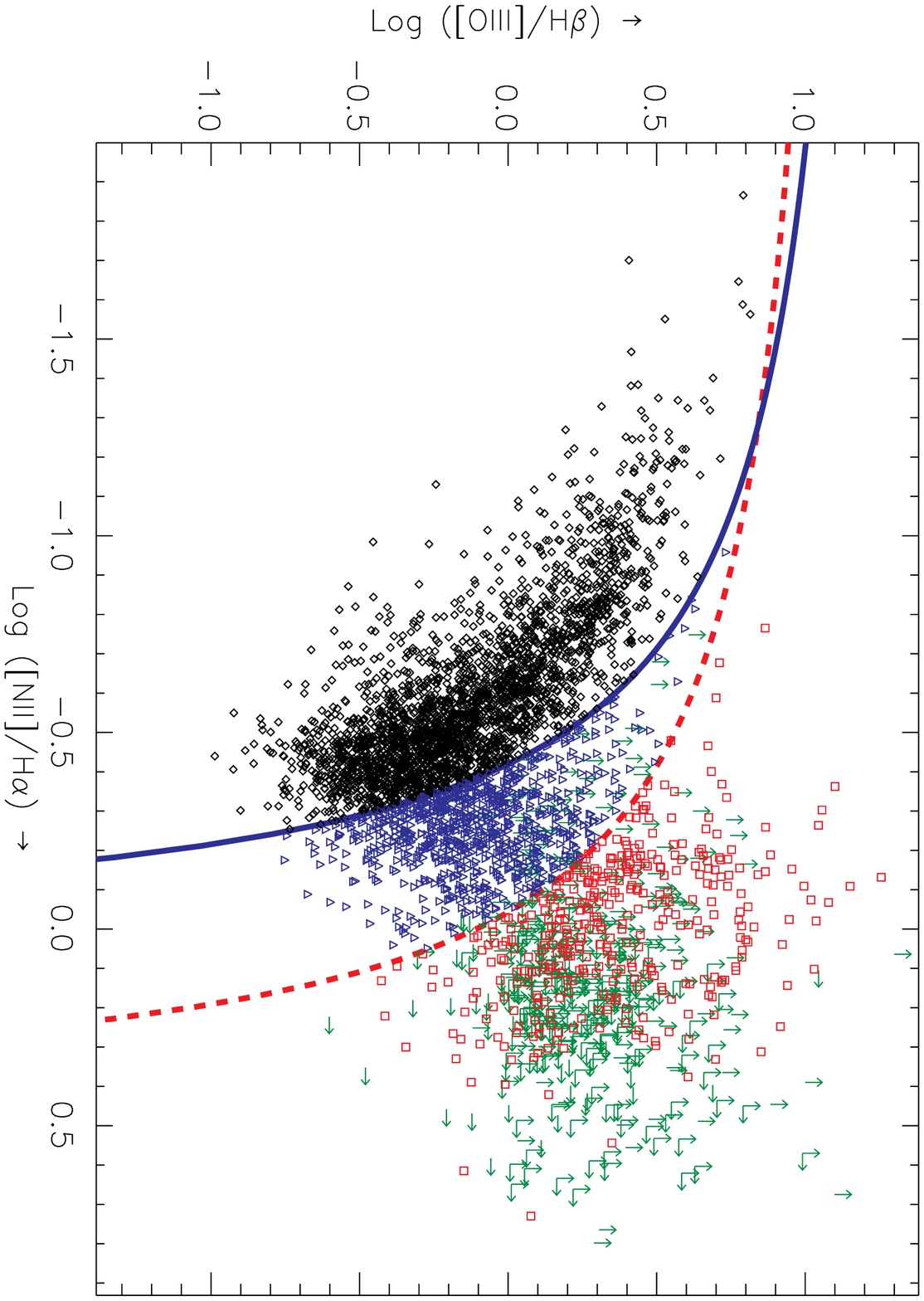}
  \caption{BPT \citep[after][]{Baldwin81} diagram for SHELS. The solid
  blue and dashed red lines indicate the demarcation of pure star
  formation from \citet{Kauffmann03} and of extreme starbursts from
  \citet{Kewley01}. We classify the galaxies as: pure star forming
  galaxies ({\it black diamonds}), AGNs ({\it red squares}), and
  composite galaxies ({\it blue triangles}). We indicate galaxies or
  AGNs with either \ha{} or \hb{} undetected, i.e. $\mathrm{S/N} < 3$,
  as lower limits ({\it green arrows}).}
  \label{fig:agn}
\end{figure}

We use the demarcations of pure star formation from
\citet{Kauffmann03} and of extreme starburst from \citet{Kewley01} to
identify galaxies as pure star forming, AGN, or a combination
(composite galaxies) based on the line ratios of \oiii{}/\hb{} and
\nii{}/\ha{} (Figure~\ref{fig:agn}).

We select all galaxies with \Oiiib{} and \Niib{} detected with a
signal-to-noise ratio (S/N) $\ge 3$. If the galaxies have both \ha{}
and \hb{} detected with S/N $\ge 3$, we identify them as pure star
forming galaxies when their line ratios are below the
\citeauthor{Kauffmann03} relation, pure AGNs when the ratios are above
the \citeauthor{Kewley01} relation, and composites when they lie
between the relations.

For galaxies with either \ha{} or \hb{} undetected (S/N $<$ 3), we use
the 3\,$\sigma$ value for the line flux to calculate the line
ratios. These ratios are lower limits. We classify these galaxies as
composite or AGN; some of the composite galaxies might be AGNs.

We identify a separate class of broad-line AGN. The width of these
broad Balmer-lines extends beyond the limited-width top-hat filter
used for measuring the line fluxes (Section~\ref{sec:emabmeasure}). In
some cases the \Nii{} lines are not distinguishable from the \ha{}
line in a spectrum with a very broad \ha{} line.

%%sigma = 5 pix = 6 A -> FWHM = 2.3548 sigma = 14.1 A
Inspecting each spectrum would be time-consuming. Hence, we fit the
\ha{} and \hb{} lines in the continuum-subtracted spectra in an
automated way and individually inspected each candidate broad-line
AGN. We fit both lines simultaneously with the assumption that the
full-width-half-maximum (FWHM) of the line profile is the same for
both lines. Candidate broad-line AGNs have a peak of both \ha{} and
\hb{} $> 5 \times \pow{-18}$\,\ergsPerAng{} above the continuum
residuals (which avoids the inclusion of noise peaks) and a FWHM of
the Gaussian component of the line profile (we use a Gaussian
convolved with the instrumental profile as our line profile) before
convolution larger than 14\,\AA{}. From these candidates, we select
the galaxies that are genuine broad-line AGNs.

The fraction of galaxies identified as AGN and/or composite over the
redshift ranges 0.010-0.100, 0.100-0.200, 0.200-0.300 and 0.300-0.377
for an \ha{} luminosity limited sample ($\log L \ge 41.18$; lowest
\ha{} luminosity at $z = 0.377$) is 5.9, 6.6, 5.3 and 5.2\,\%,
respectively.

The fraction of AGN is more or less constant with redshift. However,
we cannot draw any conclusions about the evolution of the AGN-fraction
as a function of redshift. We removed stellar objects from the initial
sample and thus may have inadvertently removed AGNs particular at
greater redshifts.

\section{The \ha{} luminosity function at redshift $\sim$ 0.24}
\label{sec:nbvbb}
The recent advent of wide-field cameras on telescopes has aided
searches for star forming galaxies by increasing the area (and hence
volume) of narrowband surveys, e.g. \citet{Fujita03}, \citet{Ly07},
\citet{Shioya08}, \citet{Westra08}, and many more. This technique has
recently been extended to the near-infrared, e.g. \citet{Sobral09}.

A narrowband survey efficiently probes the faint end of the luminosity
function which is hard to explore in a spectroscopic survey. In
contrast, a spectroscopic survey can cover a larger volume and sample
the rare luminous galaxies at the bright end of the luminosity
function.

Here, we combine the strength of a narrowband survey--the ability to
go deep--with that of our broadband selected spectroscopic
survey--coverage of a large volume--to determine a well-constrained
luminosity function at $z\sim0.24$. For the narrowband survey we use
the publicly available data from \citet[hereafter
\citetalias{Shioya08}]{Shioya08} together with that of the Cosmic
Evolution Survey \citep[COSMOS\footnote{The COSMOS catalog can be
downloaded from
\url{http://irsa.ipac.caltech.edu/data/COSMOS/tables/cosmos\_phot\_20060103.tbl.gz}};][]{Capak07}
which formed the basis of the survey of \citetalias{Shioya08}. We use
the spectroscopic survey of SHELS for the bright end of the luminosity
function.

\begin{figure*}[tb]
  \centering
  \includeIDLfigPcustom[0.75\textwidth]{20pt}{25pt}{29pt}{34pt}{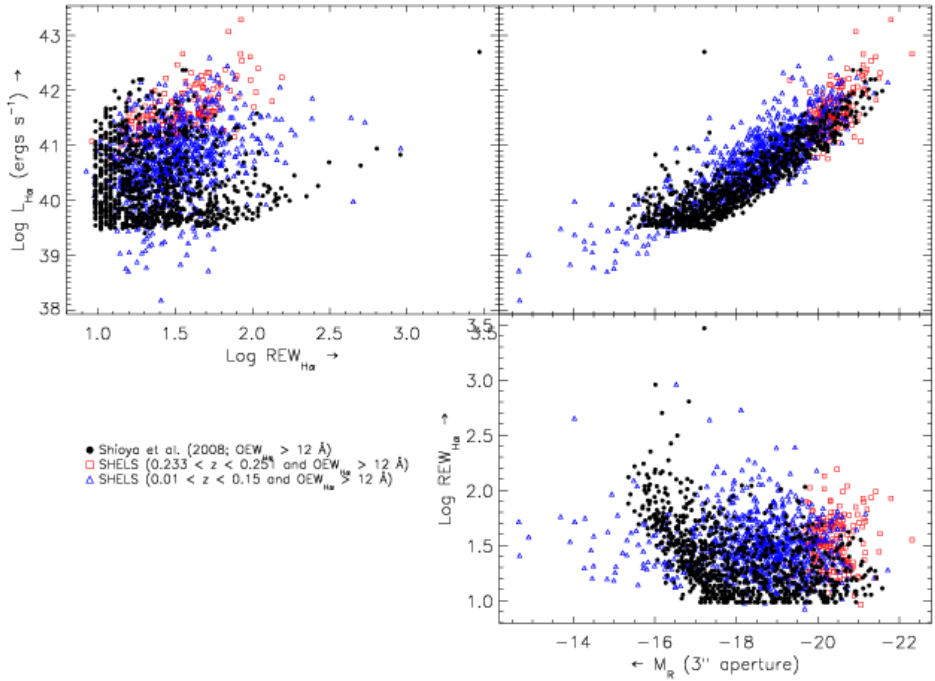}
  \caption{Comparison of SHELS with \citetalias{Shioya08} for galaxies
    with the OEW$_{\ha+\nii} > 12$\,\AA{} not corrected for \nii{} for
    both surveys. The data are: \citetalias{Shioya08} ({\it solid
    black circles}), SHELS in the redshift range of
    \citetalias{Shioya08} ({\it open red squares}), and SHELS at $0.01
    < z < 0.15$ ({\it open blue triangles}). The \ha{} luminosity,
    $L_\ha$, for both surveys is corrected for \nii{}.}
  \label{fig:shishelscomp}
\end{figure*}

The \citetalias{Shioya08} and SHELS surveys use different approaches
(imaging versus spectroscopy). A comparison of the data allows a
consistency check of the aperture corrections applied to the SHELS
data.

We construct the \ha{} luminosity function over the redshift range of
\citetalias{Shioya08} ($0.233 < z < 0.251$) based on the catalog with
emission-line fluxes determined in Section~\ref{sec:emabmeasure}
(which already include corrections for underlying stellar absorption),
redshifts, extinction corrections from Section~\ref{sec:extcorr}, and
removal of composites and AGNs (Section~\ref{sec:agn}). Constraining
SHELS to the same redshift range yields a sample of 192 SHELS galaxies
at $0.233 < z < 0.251$.

\subsection{Data comparison}
\label{sec:comparison}
Figure~\ref{fig:shishelscomp} shows the \ha{} luminosity, \ha{}
rest-frame equivalent width (REW), and the 3\arcsec{} aperture
absolute $R$-band magnitude from \citetalias{Shioya08} ({\it solid
black circles}) matched to the selection criteria of SHELS,
$R_\mathrm{tot} \leq 20.3$. We also show the SHELS data ({\it open red
squares}) for the redshift range covered by \citetalias{Shioya08}
($0.233 < z < 0.251$). To match \citetalias{Shioya08} we require an
observed equivalent width of \ha{} combined with \nii{} $\ge
12$\,\AA{} (as per the selection criteria of \citetalias{Shioya08}),
and $f_\mathrm{\ha} \ge \pow{-15.5}$\,\fluxunits{}. The strengths of
both surveys are immediately apparent. SHELS includes the highest
luminous galaxies, \citetalias{Shioya08} probes the faint end of the
luminosity function.

\begin{figure}[tb]
  \centering
  \includeIDLfigPcustom{6pt}{16pt}{12pt}{8pt}{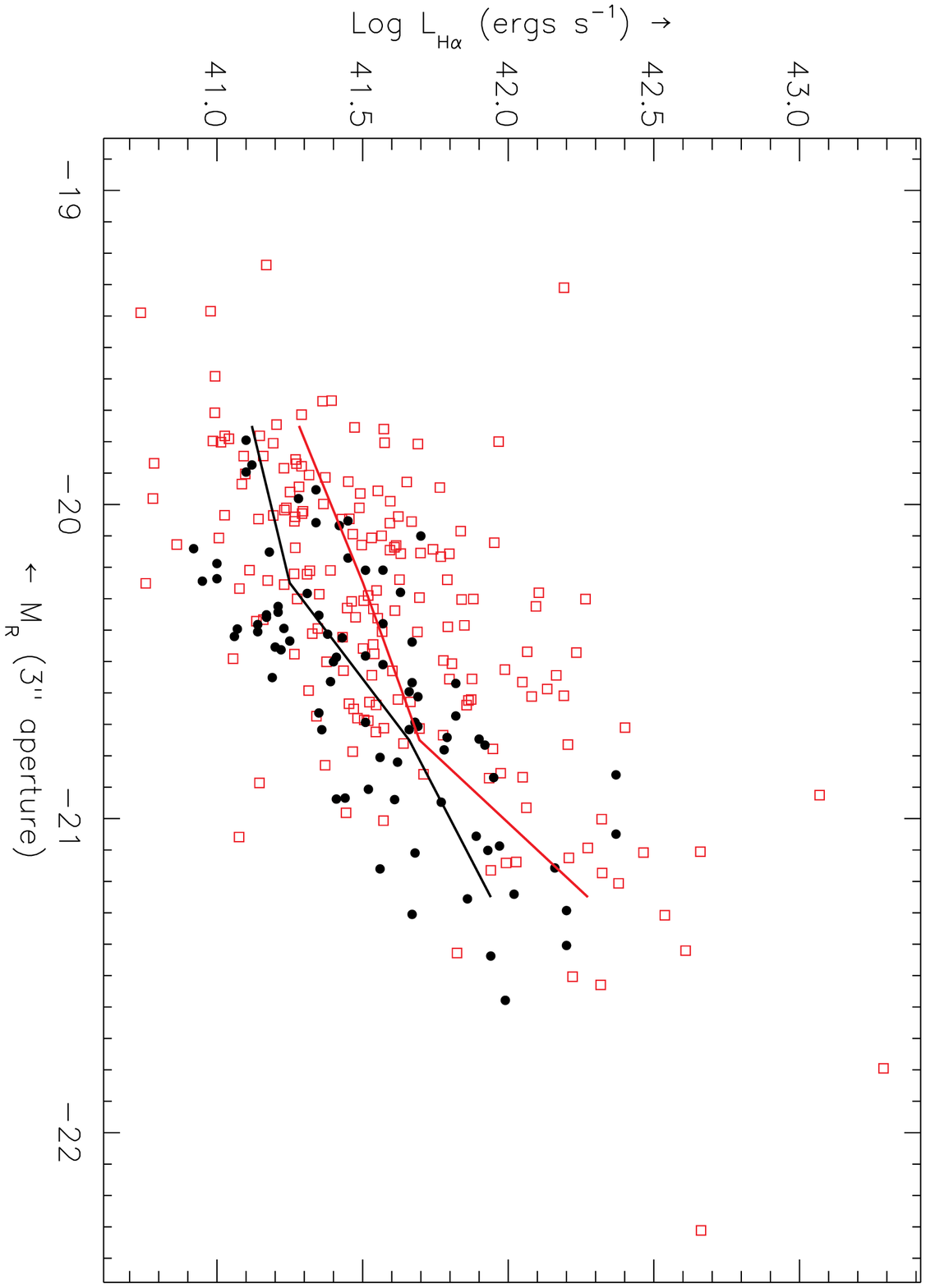}
  \caption{Comparison of the total \ha{} luminosity with the
    3\arcsec{} aperture $R$-band magnitude of SHELS ({\it open red
    squares}) and the survey of \citetalias{Shioya08} ({\it solid
    black circles}). The galaxies have OEW$_\mathrm{\ha+\nii} \ge
    12$\,\AA{}, a flux-limit of $f_\mathrm{\ha} \ge
    \pow{-15.5}$\,\fluxunits{}, $R_\mathrm{tot} \leq 20.3$, and $0.233
    < z < 0.251$. We show the median \ha{} luminosity for 0.5
    magnitude-wide bins for SHELS ({\it red line}) and
    \citetalias{Shioya08} ({\it black line}). The SHELS galaxies shift
    toward greater $L_\mathrm{\ha}$ at fixed $M_R$}
  \label{fig:shishelsRHa}
\end{figure}

\begin{figure}[tb]
  \centering
  \includeIDLfigPcustom{16pt}{8pt}{25pt}{15pt}{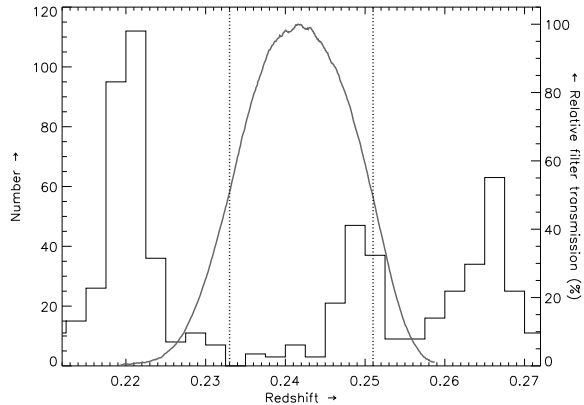}
  \caption{Redshift distribution of the galaxies in zCOSMOS DR2
    \citep[][{\it histogram}]{Lilly07}, the relative NB816 filter
    transmission curve used by \citetalias{Shioya08} ({\it gray solid
      line}), and where the transmission of the NB816 filter is 50\,\% of
    its maximum ({\it dotted lines}).}
  \label{fig:zcosmos}
\end{figure}

\citetalias{Shioya08} measure magnitudes from a 3\arcsec{} aperture,
scale them to the total $i'$-band magnitude, and calculate \ha{}
fluxes. We measure fluxes in a similar way. We use spectra taken with
a 1\farcs{}5-fiber aperture scaled to the total $R$-band
magnitude. The data from the lower redshift range $0.01 < z < 0.15$ of
SHELS (Figure~\ref{fig:shishelscomp}; {\it open blue triangles}) show
that the relation between the \ha{} luminosity and the 3\arcsec{}
total $R$-band magnitude of the two surveys is similar. The scaling of
the \ha{} flux from the limited-aperture magnitude to the total
magnitude introduces no systematic biases and is consistent with
\citetalias{Shioya08}.

If we constrain the data of \citetalias{Shioya08} to $f_\mathrm{\ha} =
\pow{-15.5}$\,\fluxunits{}, a difference at the bright end becomes
apparent (see Figure~\ref{fig:shishelsRHa}). There are more-luminous
galaxies in SHELS than in \citetalias{Shioya08}. This difference
results from two effects: ({\it i}) SHELS probes a larger volume, and
({\it ii}) the \ha{} fluxes determined from narrowband surveys can
easily underestimate the true line flux. Galaxies with redshifts that
place the \ha{} line in the wings of the filter underestimate the mean
recovered \ha{} flux.

To examine the redshift distribution of the galaxies in
\citetalias{Shioya08}, Figure~\ref{fig:zcosmos} shows the redshift
distribution of the 10k zCOSMOS catalog\footnote{zCOSMOS DR2, which
can be obtained from the ESO archives.} \citep{Lilly07} in combination
with the filter transmission curve of the NB816 normalized to the
maximum throughput\footnote{The filter profile is available at
\url{http://www.naoj.org/Observing/Instruments/SCam/txt/NB816.txt}.}. Any
galaxy with a redshift placing it in the wings of the narrowband
filter has its \ha{} flux underestimated far more than the 21\,\%
\citetalias{Shioya08} use to correct their line fluxes. In the COSMOS
field the galaxies tend to be at redshifts towards the red edge of the
filter. In this case, the \Niib{} line (the strongest of the two
\nii{} lines that straddle \ha{}) barely contributes to the flux
probed by the filter. Both the underestimation of the \ha{} flux and
over-correction for \nii{} can explain the difference in the
distribution of \ha{} fluxes in Figure~\ref{fig:shishelsRHa}.

Despite this difference, we can still use the fainter galaxies from
\citetalias{Shioya08} to determine the faint-end slope of the \ha{}
luminosity function. The systematic underestimation of fluxes causes a
shift in the luminosity function which affects the determination of
the characteristic luminosity (i.e. bright end), not the faint-end
slope.

\subsection{Derivation and fit}
\label{sec:combinedlf}

We fit a Schechter function \citep{Schechter76} to the SHELS and
\citetalias{Shioya08} data. The Schechter function is
\begin{equation}
\phi(L) dL = \phi^* \left ( \frac{L}{L^*} \right ) ^{-\alpha} \exp
\left ( -\frac{L}{L^*} \right ) d \left ( \frac{L}{L^*} \right ),
\end{equation}
where $\alpha$ is slope of the faint-end part, $L^*$ is a
characteristic luminosity, and $\phi^*$ is the
normalization. Throughout this paper the units for the Schechter
parameters $L^*$ and $\phi^*$ are \ergs{} and \perMpcQ{},
respectively. $\alpha$ is dimensionless.

To combine the two data sets we use the non-parametric
1/V$_\mathrm{max}$ method \citep{Schmidt68} to determine the Schechter
parameters. The number density of galaxies for each luminosity bin $j$
with a width of $\Delta \log L$ is
\begin{equation}
\phi(L_j) \Delta \log L = \sum \limits _{i=1} ^{N_\mathrm{gal}}
\frac{W(L_i)}{V_i},
\end{equation}
where $W(x) = 1$ when the luminosity is enclosed by bin $j$ and $W(x)
= 0$ otherwise, and $V_i$ is the volume sampled by galaxy $i$. The
uncertainties in the bins are Poisson errors
\begin{equation}
\sigma^2_{\phi(L_j)} = \sum \limits _{i=1} ^{N_\mathrm{gal}}
\frac{W(L_i)}{V_i}.
\end{equation}

\begin{figure}[tb]
  \centering
  \includeIDLfigPcustom{3pt}{28pt}{29pt}{5pt}{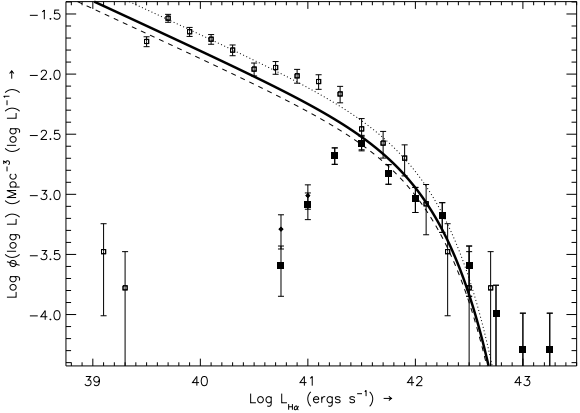}
  \caption{The 1/V$_\mathrm{max}$ data-points for SHELS ({\it solid
      squares}), SHELS without the OEW$_\mathrm{\ha+\nii} \ge
      12$\,\AA{} criterion ({\it solid diamonds}), and
      \citetalias{Shioya08} ({\it open squares}). The galaxies have
      OEW$_\mathrm{\ha+\nii} \ge 12$\,\AA{}, $R_\mathrm{tot} \leq
      20.3$, and $0.233 < z < 0.251$. The {\it thick solid line}
      indicates the combined fit of SHELS and \citetalias{Shioya08}
      with $\phi^*_\mathrm{comb} = -3.05 \pm 0.09$, $\alpha = -1.41
      \pm 0.03$, and $\log L^* = 42.14 \pm 0.08$. The {\it thin lines}
      indicate the luminosity function for the SHELS and
      \citetalias{Shioya08} data separately with $\phi ^*
      _\mathrm{SHELS} = -3.11 \pm 0.09$ ({\it dashed line}) and $\phi
      ^* _\mathrm{S08} = -2.91 \pm 0.09$ ({\it dotted line}),
      respectively. Both luminosity functions also have $\alpha =
      -1.41 \pm 0.03$ and $\log L^* = 42.14 \pm 0.08$.}
  \label{fig:combinedLF}
\end{figure}

Figure~\ref{fig:combinedLF} shows the luminosity function for the
combined data set ({\it thick solid line}). For both surveys we apply
the selection criteria $R_\mathrm{tot} \leq 20.3$,
OEW$_\mathrm{\ha+\nii} \ge 12$\,\AA{}, and $0.233 < z < 0.251$.

We determine the data-points using 1/V$_\mathrm{max}$ where the
uncertainties are Poisson errors for both SHELS ({\it solid squares})
and \citetalias{Shioya08} ({\it open squares}). We fit a Schechter
function with common $L^*$ and $\alpha$ to the combined SHELS and
\citetalias{Shioya08} dataset. For this fit, we use the data with
$\log L_\ha{} \ge 41.4$ for SHELS and $\log L_\ha{} \ge 39.6$ for
\citetalias{Shioya08}. We recover a single value for $\alpha$ and
$L^*$ of the joint fit: $\alpha = -1.41 \pm 0.03$ and $\log L^* =
42.14 \pm 0.08$. For those values, the normalization for SHELS is
$\log \phi ^* _\mathrm{SHELS} = -3.11 \pm 0.09$ and for
\citetalias{Shioya08} is $\log \phi ^* _\mathrm{S08} = -2.91 \pm
0.09$. We combine $\phi ^* _\mathrm{S08}$ and $\phi ^*
_\mathrm{SHELS}$ using a volume-weighted average. Thus, $\log \phi ^*
_\mathrm{comb} = \log[(1.5 \times \pow{-2.91} + 4.0 \times
\pow{-3.11})/5.5] = -3.05$. This method of combining is one way to
account for cosmic variance. We adopt $\alpha = -1.41 \pm 0.03$, $\log
L^* = 42.14 \pm 0.08$, $\log \phi ^* _\mathrm{comb} = -3.05 \pm 0.09$
for comparison to other surveys.

\section{The \ha{} luminosity functions from SHELS}
\label{sec:halfSHELS}

\begin{deluxetable}{llccccc}
  \tablewidth{0pt}

  \tablecaption{Number of galaxies satisfying each selection
  criterion.\label{tab:ngals}}

  \tablehead{
    \multicolumn{2}{c}{Criterion} & \colhead{$0.01 \le z < 0.10$} & \colhead{$0.10 \le z < 0.20$} & \colhead{$0.20 \le z < 0.30$} & \colhead{$0.30 \le z < 0.38$} & \colhead{Total}
  }

  \startdata
  ({\it i})   & $R_\mathrm{tot} \leq 20.3$      &  461 & 1949 & 2746 & 2114 & 7270\\
  ({\it ii})  & $\mathrm{S/N}_\mathrm{\ha} > 5$ &  420 & 1640 & 1857 & 1275 & 5192\\
  ({\it iii}) & $\log _\mathrm{\ha} \ge -15.5$  &  369 & 1441 & 1702 & 1186 & 4698\\
  ({\it iv})  & pure star forming               &  322 & 1127 & 1268 & \phn848 & 3565\\			 			    			    										  			  					  			  						  			  
  \enddata

  \tablecomments{Successive lines are a subset of the line above. We
  apply the selection criteria sequentially in the order of the
  Table.}

\end{deluxetable}

Here we use SHELS to determine the \ha{} luminosity functions as a
function of redshift. We can identify \ha{} in our spectra up to a
redshift of $z_\mathrm{max} = 0.377$. We next examine the influence of
the $R$-magnitude limited survey on the derivation of the \ha{}
luminosity function. We limit our sample to $f_\mathrm{\ha} \ge
\pow{-15.5}$\,\fluxunits{} (see
Section~\ref{sec:sampledef}). Table~\ref{tab:ngals} lists the number
of galaxies satisfying each selection
criterion. Figure~\ref{fig:lhavz} shows the distribution of the \ha{}
luminosity as a function of redshift and the redshift bins used to
construct the \ha{} luminosity functions.

\subsection{Derivation and fit}
\label{sec:styfit}
Here we use the STY-method \citep{Sandage79}, a parametric estimation
method, to determine the three Schechter parameters for each
redshift bin. The STY-method identifies the luminosity function
parameters that maximize the probability of obtaining the observed
sample. The probability $\mathcal{P}$ is
\begin{equation}
\mathcal{P} = \prod \limits _{i=1} ^{N_\mathrm{gal}}
\frac{\phi(L_i)}{\int ^\infty _{L(z_i)_{\mathrm{lim},i}} \phi(L) dL},
\end{equation}
where $L(z)_{\mathrm{lim},i}$ is the faintest luminosity where galaxy
$i$ at redshift $z_i$ is observable. We use a truncated-Newton method
to maximize the natural logarithm of $\mathcal{P}$.

\begin{figure*}[tb]
  \centering
  \includeIDLfigP[0.75\textwidth]{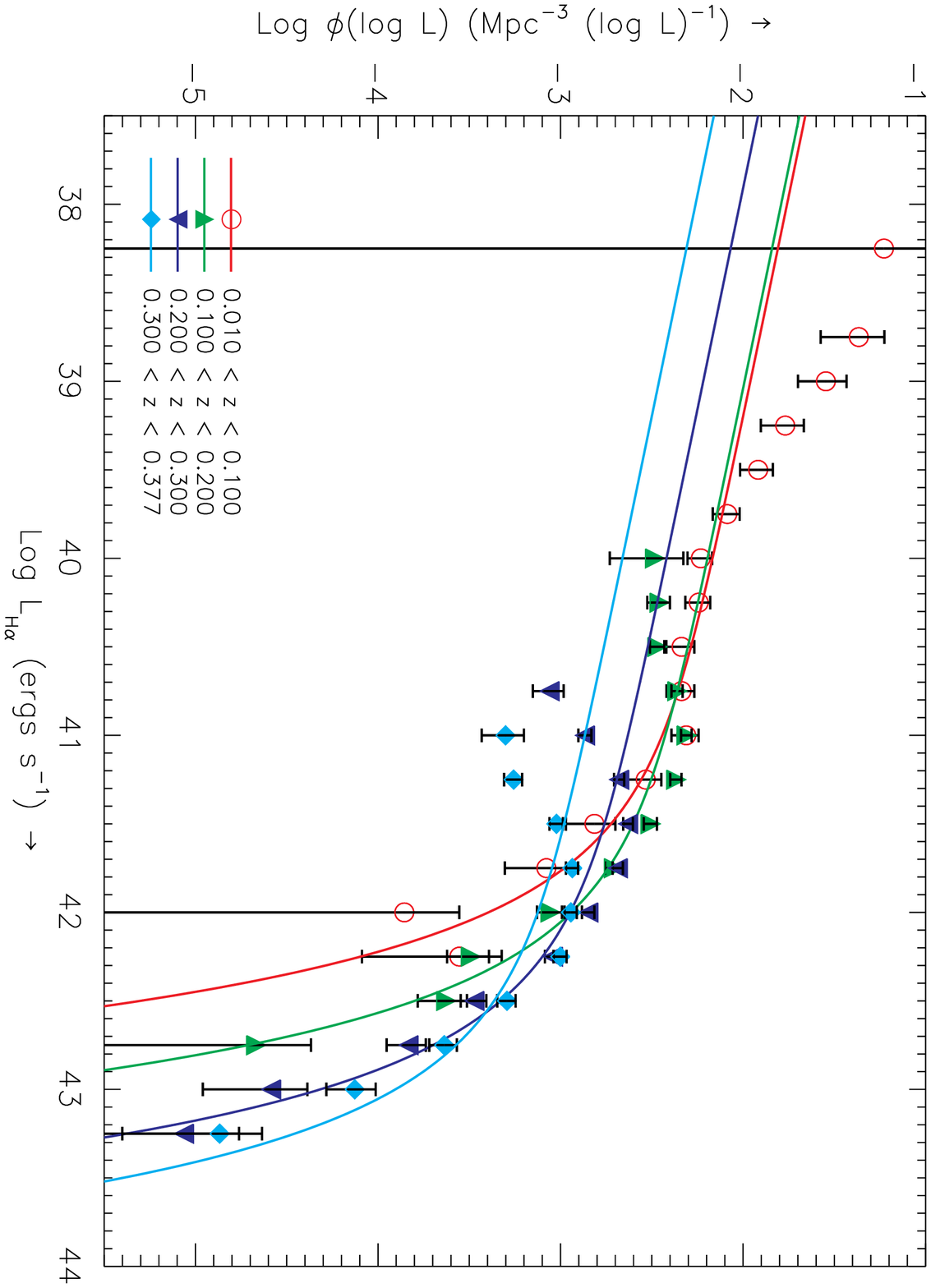}
  \caption{\ha{} luminosity functions for several redshift bins for
    our pure star forming sample. The Schechter functions are derived
    using the STY-method with $\alpha$ fixed to $-1.2$. The
    data-points come from 1/V$_\mathrm{max}$ where the uncertainties
    are Poisson errors. Table~\ref{tab:schechter} lists the
    parameters.}
  \label{fig:schechter}
\end{figure*}

Table~\ref{tab:schechter} lists the fit-parameters, and
Figure~\ref{fig:schechter} shows the results. We fit for $\alpha$,
$L^*$, and $\phi^*$ ({\it dashed lines}). We also use a fixed $\alpha
= -1.20$ ({\it solid lines}). This fixed value represents the slope
over the redshift range $0.05 < z < 0.20$ for SHELS. This range has a
large enough volume to sample the bright end of the luminosity
function, while still having galaxies faint enough to determine the
faint end slope. We do not consider this large redshift range in our
further analysis.

Narrowband surveys apply a correction to the total luminosity density
for galaxies hosting an AGN
\citep[e.g.][]{Fujita03,Ly07,Shioya08,Westra08}. A large fraction of
these surveys have little or no spectroscopy. Thus there is no way to
separate AGNs from star forming galaxies. SHELS enables a direct
separation (see Section~\ref{sec:agn}). We derive the Schechter
parameters for the \ha{} emitting galaxies with the AGNs removed
(Table~\ref{tab:schechter}). Removal of AGNs moves $L^*$ slightly
fainter and reduces the normalization because AGNs are systematically
in more luminous galaxies. Failure to account for this bias introduces
a systematic offset.

\begin{deluxetable}{lcccccc}
  \tablewidth{0pt}

  \tablecaption{Parameters for the \ha{} luminosity
  functions.\label{tab:schechter}}

  \tablehead{
    \colhead{} & \multicolumn{3}{c}{fixed $\alpha$} & \multicolumn{3}{c}{unconstrained $\alpha$}\\
    \colhead{redshift range} & \colhead{$\alpha$} & \colhead{$\log L^*$} & \colhead{$\log \phi^*$} & \colhead{$\alpha$} & \colhead{$\log L^*$} & \colhead{$\log \phi^*$}
  }

  \tablewidth{0pt}

  \startdata

  \multicolumn{7}{c}{Pure star forming sample}\\
  $0.010 < z < 0.100$\tablenotemark{a} & $-1.20$ & $41.72 \pm 0.10$ & $-2.86 \pm 0.04$ & $-1.22 \pm 0.06$ & $41.74 \pm 0.13$ & $-2.90 \pm 0.10$\\
  $0.100 < z < 0.200$ & $-1.20$ & $42.09 \pm 0.04$ & $-2.97 \pm 0.02$ & $-0.87 \pm 0.05$ & $41.79 \pm 0.06$ & $-2.58 \pm 0.05$\\
  $0.200 < z < 0.300$ & $-1.20$ & $42.52 \pm 0.04$ & $-3.29 \pm 0.01$ & $-0.71 \pm 0.07$ & $42.13 \pm 0.06$ & $-2.79 \pm 0.05$\\
  $0.300 < z < 0.377$ & $-1.20$ & $42.83 \pm 0.03$ & $-3.59 \pm 0.01$ & $-0.50 \pm 0.06$ & $42.30 \pm 0.05$ & $-2.96 \pm 0.04$\\
  $0.233 < z < 0.251$\tablenotemark{b} & \multicolumn{3}{c}{} & $-1.41 \pm 0.03$ & $42.14 \pm 0.08$ & $-3.05 \pm 0.09$\\

  \multicolumn{7}{c}{Pure star forming sample including composites and AGNs}\\
  $0.010 < z < 0.100$\tablenotemark{a} & $-1.20$ & $41.76 \pm 0.10$ & $-2.82 \pm 0.04$ & $-1.25 \pm 0.06$ & $41.85 \pm 0.13$ & $-2.93 \pm 0.10$\\
  $0.100 < z < 0.200$ & $-1.20$ & $42.13 \pm 0.04$ & $-2.88 \pm 0.02$ & $-0.99 \pm 0.05$ & $41.92 \pm 0.06$ & $-2.61 \pm 0.05$\\
  $0.200 < z < 0.300$ & $-1.20$ & $42.55 \pm 0.04$ & $-3.17 \pm 0.01$ & $-0.88 \pm 0.07$ & $42.26 \pm 0.06$ & $-2.80 \pm 0.05$\\
  $0.300 < z < 0.377$ & $-1.20$ & $42.83 \pm 0.03$ & $-3.44 \pm 0.01$ & $-0.66 \pm 0.06$ & $42.39 \pm 0.05$ & $-2.90 \pm 0.04$\\
  \enddata

  \tablenotetext{a}{The redshift range $0.010 < z < 0.100$ covers an
  atypical under-dense region (Section~\ref{sec:fieldselection}).}

  \tablenotetext{b}{Combined SHELS and \citetalias{Shioya08}
  result. The quoted uncertainties are the formal uncertainties of the
  fit.}

  \tablecomments{For each redshift range, we list the determined
  Schechter parameters with fixed $\alpha = -1.20$ ({\it left}) and
  $\alpha$ unconstrained ({\it right}) and their uncertainties. We
  also list the parameters for the Schechter fit for a pure star
  forming sample ({\it top}) and a sample where the AGNs and
  composites are included ({\it bottom}). The pure star forming sample
  and the sample including composites and AGNS are based on criterion
  ({\it iv}) and ({\it iii}), respectively, in
  Table~\ref{tab:ngals}. The quoted uncertainties are calculated in
  Section~\ref{sec:styfit}.}
\end{deluxetable}

\begin{figure}[tb]
  \centering
  \includeIDLfigP[\columnwidth]{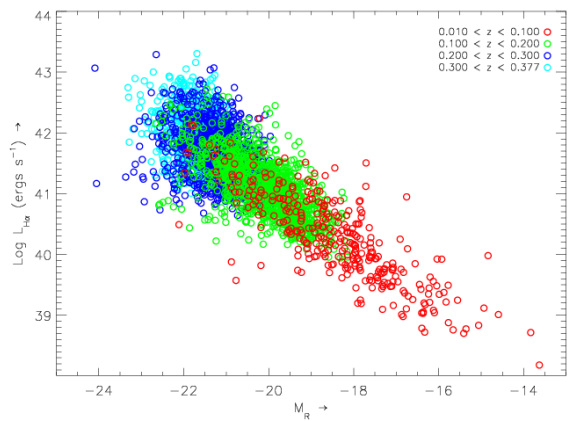}
  \caption{The distributions of $\log L_\ha$ as function of $M_R$ for
    each redshift bin. We use these distributions to assign an \ha{}
    luminosity to the observed absolute magnitude and to assign an
    absolute magnitude to a simulated \ha{} luminosity. This procedure
    enables us to determine the final uncertainties in the Schechter
    parameters and to study the influence of the survey selection
    criterion $R_\mathrm{tot} \le 20.3$.}
  \label{fig:system}
  \label{fig:propersimdist}
\end{figure}

We determine the final uncertainties in the Schechter parameters by
constructing 1,000 sets of \ha{} luminosities. We simulate the \ha{}
luminosities by converting the observed absolute magnitudes into \ha{}
luminosities using the distribution of $L_\ha{}$ as a function of
$M_R$ for each redshift bin from Figure~\ref{fig:system}. We
redetermine the Schechter parameters for each simulation using the
STY-method. The 1\,$\sigma$ spread in the redetermined parameters is
the final uncertainty which includes the formal fitting uncertainty,
uncertainties resulting from the size of the sampled volume, and the
uncertainties in the observed \ha{}
luminosity. Table~\ref{tab:schechter} lists the Schechter parameters
and their uncertainties.

\begin{figure*}[tb]
  \centering
  \includeIDLfigPcustom[0.75\textwidth]{18pt}{16pt}{15pt}{5pt}{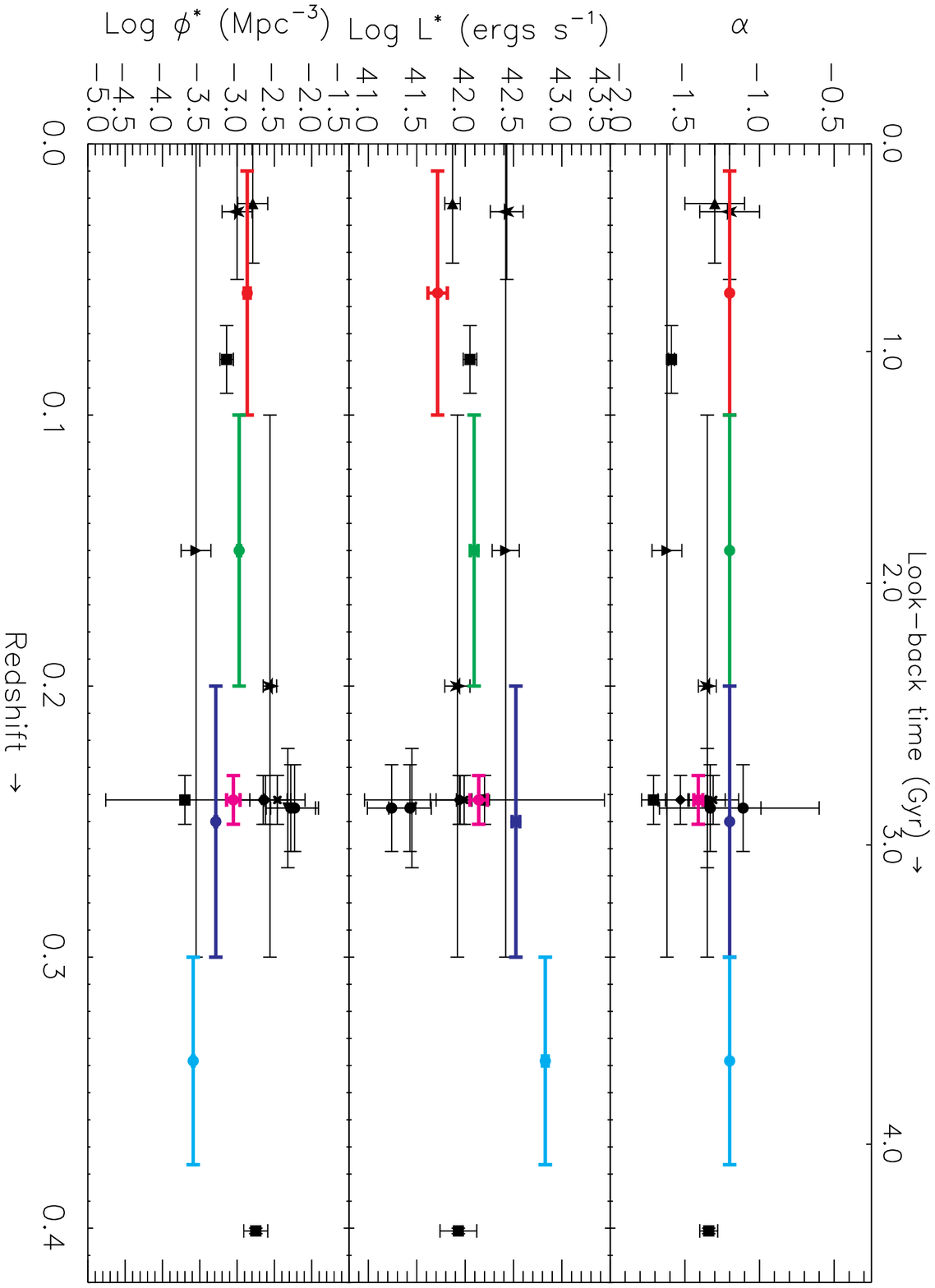}
  \includeIDLfigPcustom[0.24\textwidth]{90pt}{435pt}{40pt}{70pt}{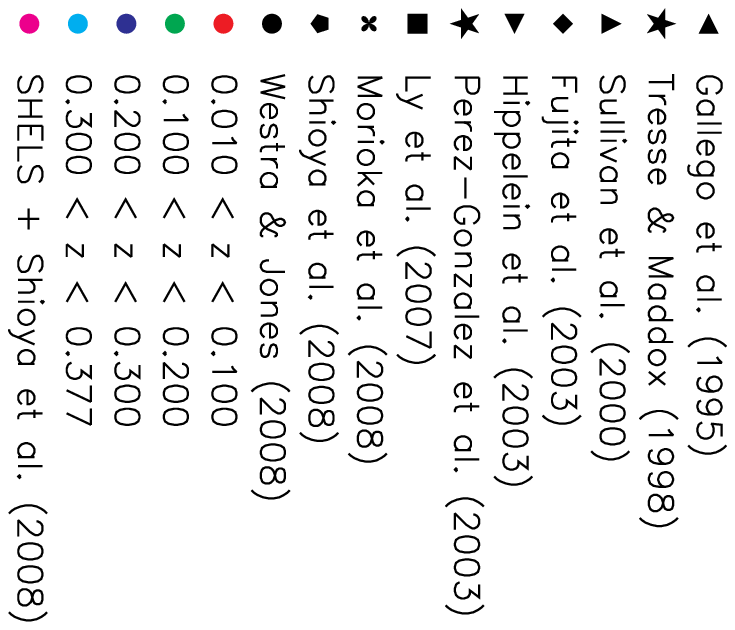}
  \caption{The three Schechter parameters as a function of redshift
  and look-back time for SHELS for pure star forming galaxies ({\it
  red, green, blue, and cyan points}), SHELS combined with
  \citetalias{Shioya08} ({\it magenta point}), and surveys at similar
  redshifts that also use \ha{} as star formation indicator ({\it
  black points}).
  }
  \label{fig:parevolution}
\end{figure*}

\subsection{Parameter evolution and impact of selection criteria}
\label{sec:parevolution}
Figure~\ref{fig:parevolution} compares the Schechter function
parameters of SHELS with fixed $\alpha = -1.20$ with other \ha{}
surveys. Evolution in the characteristic luminosity $L^*$ is clearly
visible.

The selection of galaxies by apparent $R$-band magnitude does not
yield the same sample of galaxies obtained when selecting by \ha{}
flux/luminosity. Because we determine the \ha{} luminosity function
from an $R$-selected spectroscopic survey, we must investigate the
potential systematic effects of the selection criteria
($R_\mathrm{tot} \leq 20.3$ and $f_\mathrm{\ha} \ge
\pow{-15.5}$\,\fluxunits{}) on the \ha{} luminosity function.

\begin{figure*}[tb]
  \centering
  \includeIDLfigPcustom[0.495\textwidth]{6pt}{14pt}{30pt}{5pt}{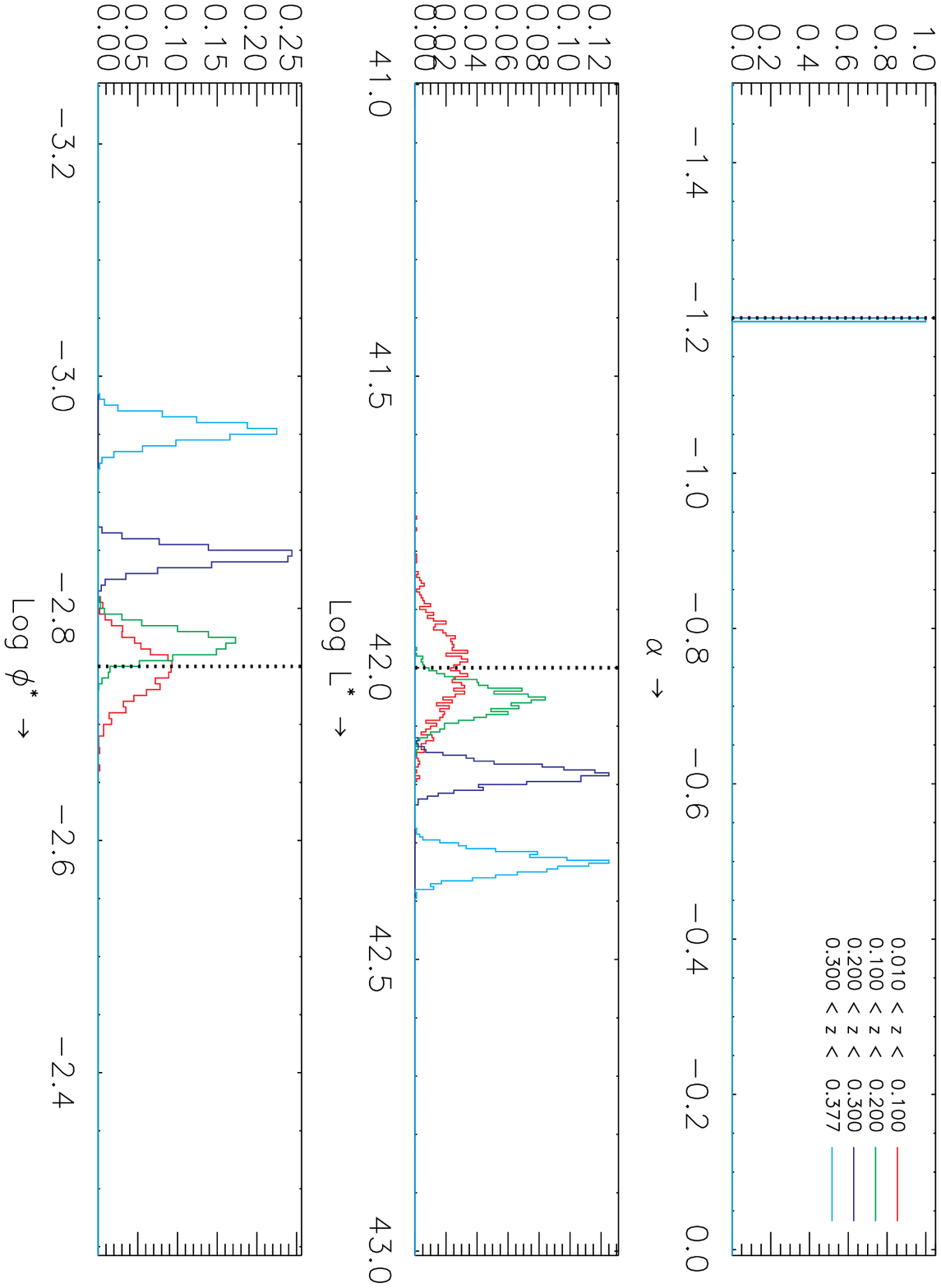}
  \includeIDLfigPcustom[0.495\textwidth]{6pt}{14pt}{30pt}{5pt}{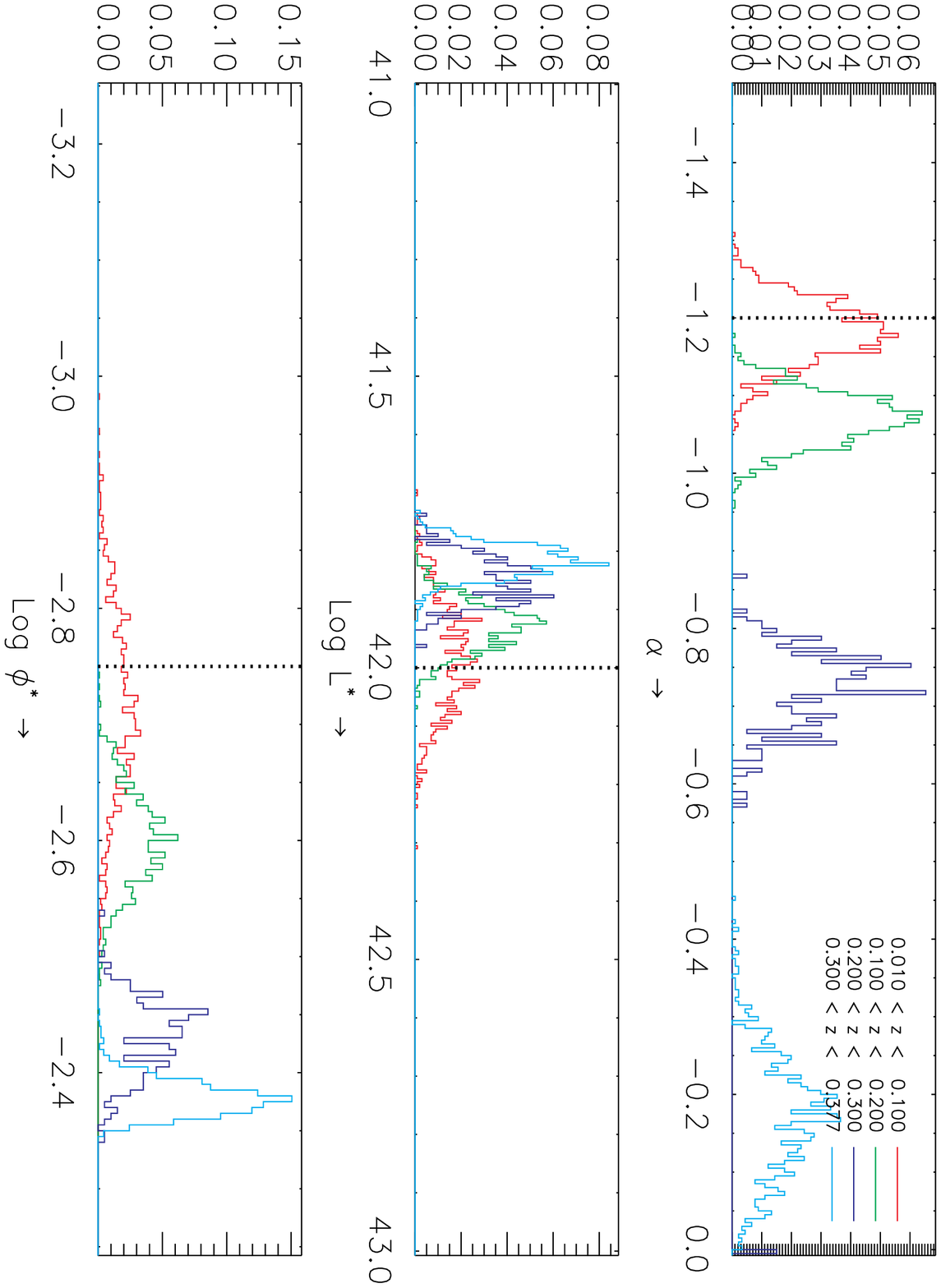}
  \caption{Histogram of the \ha{} luminosity function parameters
    determined from simulations to test the influence of the $R$-band
    selection criteria on the \ha{} luminosity function. Each
    histogram ({\it red, green, blue, cyan}) indicates the recovered
    parameters for a different redshift bin. Vertical dotted lines
    indicate the input parameters of the luminosity function to the
    simulations ($\alpha = -1.20$, $\log L^* = 42.00$, and $\log
    \phi^* = -2.75$). We indicate the results with $\alpha$ fixed at
    $-1.20$ ({\it left}) and with $\alpha$ as a free parameter ({\it
    right}). Table~\ref{tab:propersim} also shows the results.}
  \label{fig:propersimresult}
\end{figure*}

From a given luminosity function ($\alpha = -1.20$, $\log L^* (\ergs)
= 42.00$, and $\log \phi^* (\perMpcQ) = -2.75$) we construct a sample
of galaxies with a flux $f_\mathrm{\ha} \ge
\pow{-15.5}$\,\fluxunits{}. We choose these parameters because they
are close to our recovered parameters from SHELS. Furthermore, we keep
the parameters constant over our redshift range to test whether our
selection criteria introduce an artificial evolution to the
parameters. We assume a uniform galaxy distribution in a comoving
volume. We assign each simulated galaxy an absolute magnitude using
the distribution of absolute magnitude as function of $L_\mathrm{\ha}$
in Figure~\ref{fig:propersimdist}. We calculate the apparent magnitude
using the allocated redshift and a $k+e$-correction based on the
observed distribution as a function of redshift. We apply the survey
selection criterion of $R_\mathrm{tot} \leq 20.3$ and recover the
Schechter parameters using the
STY-method. Figure~\ref{fig:propersimresult} and
Table~\ref{tab:propersim} give the results.

\begin{deluxetable}{lcccccc}
  \tablewidth{0pt}

  \tablecaption{Recovered Schechter parameters for simulated \ha{}
  luminosity functions.\label{tab:propersim}}

  \tablehead{
    \colhead{} & \multicolumn{3}{c}{fixed $\alpha$} & \multicolumn{3}{c}{unconstrained $\alpha$}\\
    \colhead{redshift range} & \colhead{$\alpha$} & \colhead{$\log L^*$} & \colhead{$\log \phi^*$} & \colhead{$\alpha$} & \colhead{$\log L^*$} & \colhead{$\log \phi^*$}
  }

  \startdata

  input parameters & $-1.20$ & $42.00$ & $-2.75$ & $-1.20$ & $42.00$ & $-2.75$\\
  $0.010 \le z < 0.100$ & $-1.20$ & $42.01 \pm 0.07$ & $-2.75 \pm 0.02$ & $-1.18 \pm 0.04$ & $41.98 \pm 0.10$ & $-2.71 \pm 0.08$\\
  $0.100 \le z < 0.200$ & $-1.20$ & $42.06 \pm 0.03$ & $-2.77 \pm 0.01$ & $-1.07 \pm 0.03$ & $41.93 \pm 0.04$ & $-2.60 \pm 0.04$\\
  $0.200 \le z < 0.300$ & $-1.20$ & $42.18 \pm 0.02$ & $-2.84 \pm 0.01$ & $-0.73 \pm 0.05$ & $41.85 \pm 0.03$ & $-2.43 \pm 0.03$\\
  $0.300 \le z < 0.377$ & $-1.20$ & $42.33 \pm 0.02$ & $-2.95 \pm 0.01$ & $-0.17 \pm 0.07$ & $41.81 \pm 0.03$ & $-2.38 \pm 0.01$\\

  \enddata

\end{deluxetable}

When we keep $\alpha$ fixed in fitting the simulations, there is an
artificial trend of increasing $L^*$ and decreasing $\phi^*$ with
increasing redshift. This trend results from selective removal of the
fainter \ha{} galaxies. The removal would otherwise result in $\alpha$
decreasing (which we also show for $\alpha$ unconstrained). Thus,
$L^*$ and $\phi^*$ should be corrected for the fact that $\alpha$ is
kept fixed at a steeper value than would be fit. However, the
simulated trend in $L^*$ and $\phi^*$ is far smaller ($\Delta \log L^*
= 0.33$, $\Delta \log \phi^* = -0.20$) than we determine from the
observations ($\Delta \log L^* = 1.11$, $\Delta \log \phi^* =
-0.73$). Thus, the evolution in $L^*$ in Figure~\ref{fig:parevolution}
is real.

When we fit the simulations for all three parameters (unconstrained
$\alpha$ in Table~\ref{tab:propersim}), the decrease of $\alpha$
($\Delta \alpha = -1.01$) with increasing redshift is close to that of
the observations ($\Delta \alpha = -0.72$;
Table~\ref{tab:schechter}). Moreover, the faint-end slope from our
lowest redshift bin ($0.010 < z < 0.100$; where the faint-end of the
luminosity function is well-sampled) is consistent with that of the
combined luminosity function of SHELS and \citetalias{Shioya08} at $z
\sim 0.24$ (Section~\ref{sec:combinedlf}) within the
uncertainties. Hence, we have {\it no} evidence for evolution of the
faint-end slope over the redshift range covered by SHELS. The trend
observed with the faint-end slope unconstrained is the result of our
selection criteria.

We also notice an artificial trend in $L^*$ with increasing redshift
for a constant luminosity function (although smaller than with
$\alpha$ constrained) opposite to the trend observed, and opposite to
the trend we derive fitting for $\alpha = -1.20$. To compensate for a
shallow faint-end slope and a slight decrease in $L^*$ with redshift,
$\phi^*$ increases in these simulations.

We do not consider $\phi^*$ because it only normalizes the luminosity
function and does not determine the shape of it, unlike $\alpha$ and
$L^*$. The normalization is dependent on the number of galaxies
sampled. Because this number is heavily influenced by the distribution
of galaxies (i.e. the large-scale structure, see
Figure~\ref{fig:lhavz}), it is not possible to say anything meaningful
about any trend in $\phi^*$ even with the area covered by SHELS.

In summary, there is strong evidence for evolution in $L^*$ and no
evidence for evolution in $\alpha$ over $0.100 < z < 0.377$.

\begin{figure*}[tb]
  \centering
  \includeIDLfigPcustom[0.75\textwidth]{6pt}{14pt}{18pt}{9pt}{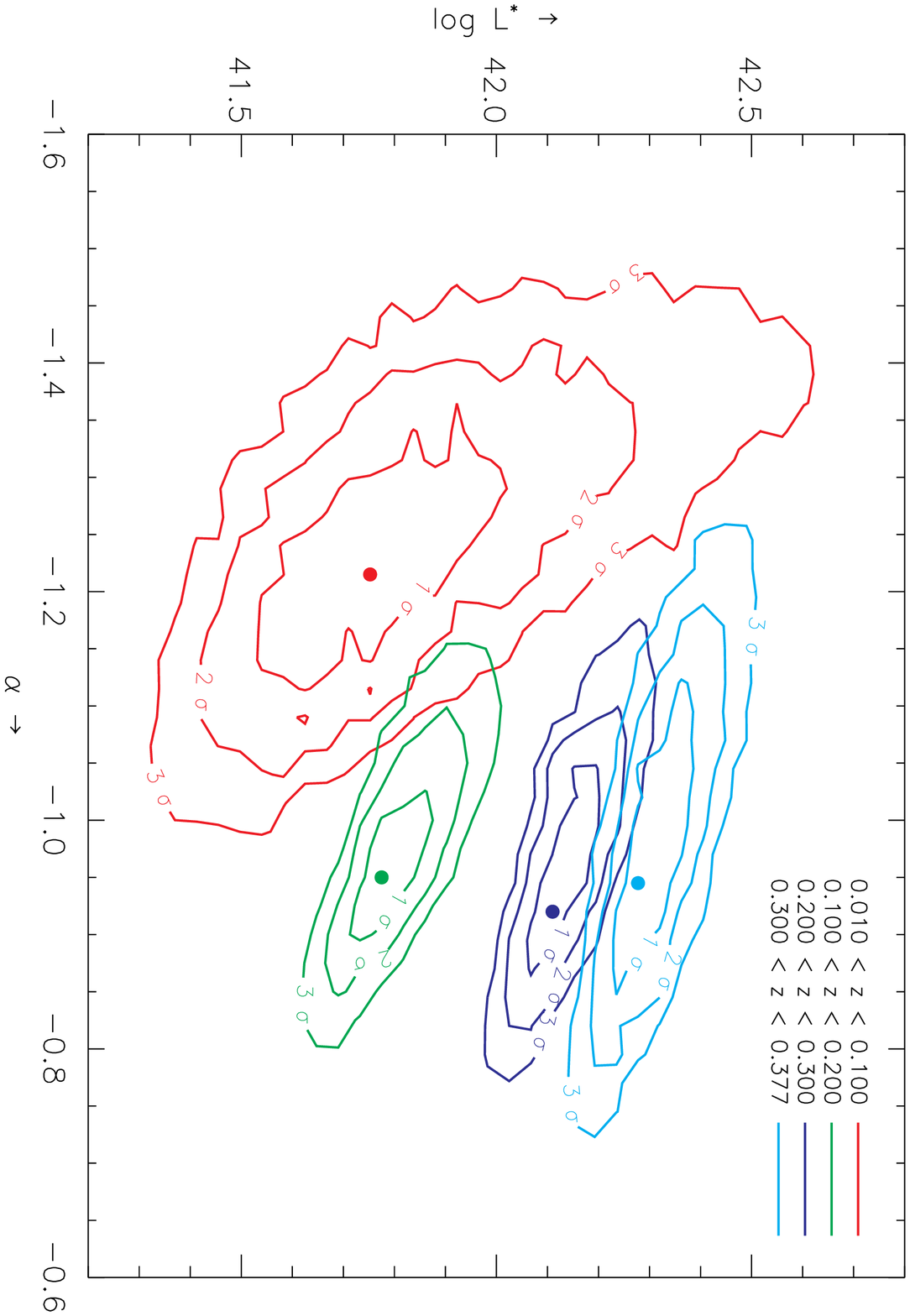}
  \caption{The luminosity function parameters for the input \ha{}
  luminosity function used to determine the observed \ha{} luminosity
  functions as shown in Figure~\ref{fig:schechter}. The contours show
  the 68.3, 95.4, and 99.7\,\% confidence intervals based on the fit
  of the output \ha{} luminosity function.}
  \label{fig:truehalf}
\end{figure*}

\subsection{The ``true'' \ha{} luminosity function}
\label{sec:truehalf}

In Section~\ref{sec:parevolution} we investigate the influence of our
selection criteria on an assumed luminosity function. We can extend
this application to determine the ``true'' \ha{} luminosity
function.

We construct a sample of galaxies with a flux $f_\mathrm{\ha} \ge
\pow{-15.5}$\,\fluxunits{} for a grid of given values of $\alpha$ and
$L^*$. We constrain $\phi^*$ by the number of observed galaxies in
each redshift bin. These choices are our input Schechter
parameters. We apply our magnitude selection of $R_\mathrm{tot} \le
20.3$. Then we determine the output: the parameters one would recover
using the STY-method. We take the median of the recovered Schechter
parameters as our final output Schechter parameters. With these final
output parameters we determine the likelihood for our observations
(Figure~\ref{fig:schechter}). We also show the input parameters and
the confidence intervals from the likelihood-determination for each
redshift bin (Figure~\ref{fig:truehalf}).

Again, we find a significant evolution of $L^*$. $L^*$ increased
towards higher redshifts, regardless of the inclusion of the lowest
redshift results. Furthermore, there is no significant evolution in
the faint end slope of the intrinsic \ha{} luminosity function. These
results confirm the findings in the
Section~\ref{sec:parevolution}. The evolution in $L^*$ is real and
there is no evidence for evolution in $\alpha$.

\begin{figure}[tb]
  \centering
  \includeIDLfigPcustom{10pt}{0pt}{26pt}{7pt}{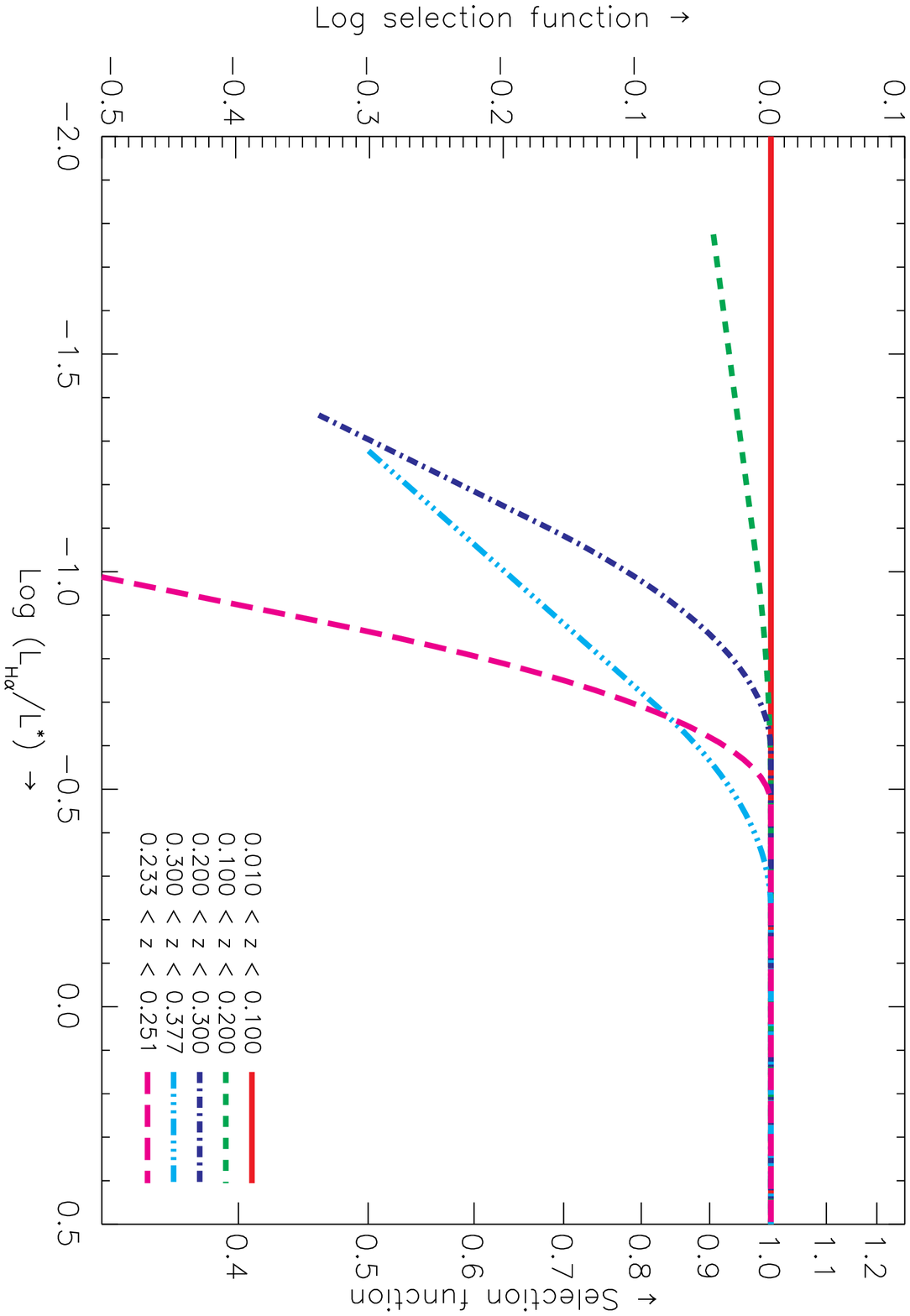}
  \caption{Logarithm of the selection function for each of the
  redshift bins sampled by SHELS as a function of \ha{} with respect
  to $L^*$ at the respective redshift bin.}
  \label{fig:haselfie}
\end{figure}

Given the input \ha{} luminosity function, we can calculate the
selection function for each redshift bin. We define the selection
function as the ratio of the measured data points in
Figure~\ref{fig:schechter} and the intrinsic or ``true'' \ha{}
luminosity function. The selection function measures the effect of our
$R \le 20.3$ selection criterion. We show the selection functions in
Figure~\ref{fig:haselfie}.

At $z\sim0.24$ we also consider the data of \citetalias{Shioya08}
These data should be complete over the luminosity range covered by
SHELS. We thus assume the data from the narrowband survey as the
intrinsic \ha{} luminosity function and take the ratio between the
\citetalias{Shioya08} and the SHELS data as an estimate of the
selection function (Figure~\ref{fig:haselfie}; {\it magenta
long-dashed line}). For consistency with the other redshift bins of
SHELS we remove the OEW$_\mathrm{\ha+\nii} \ge 12$\,\AA{}
constraint in this calculation ({\it solid diamonds} in
Figure~\ref{fig:combinedLF}).

\begin{figure}[tb]
  \centering
  \includeIDLfigPcustom[\columnwidth]{6pt}{10pt}{18pt}{9pt}{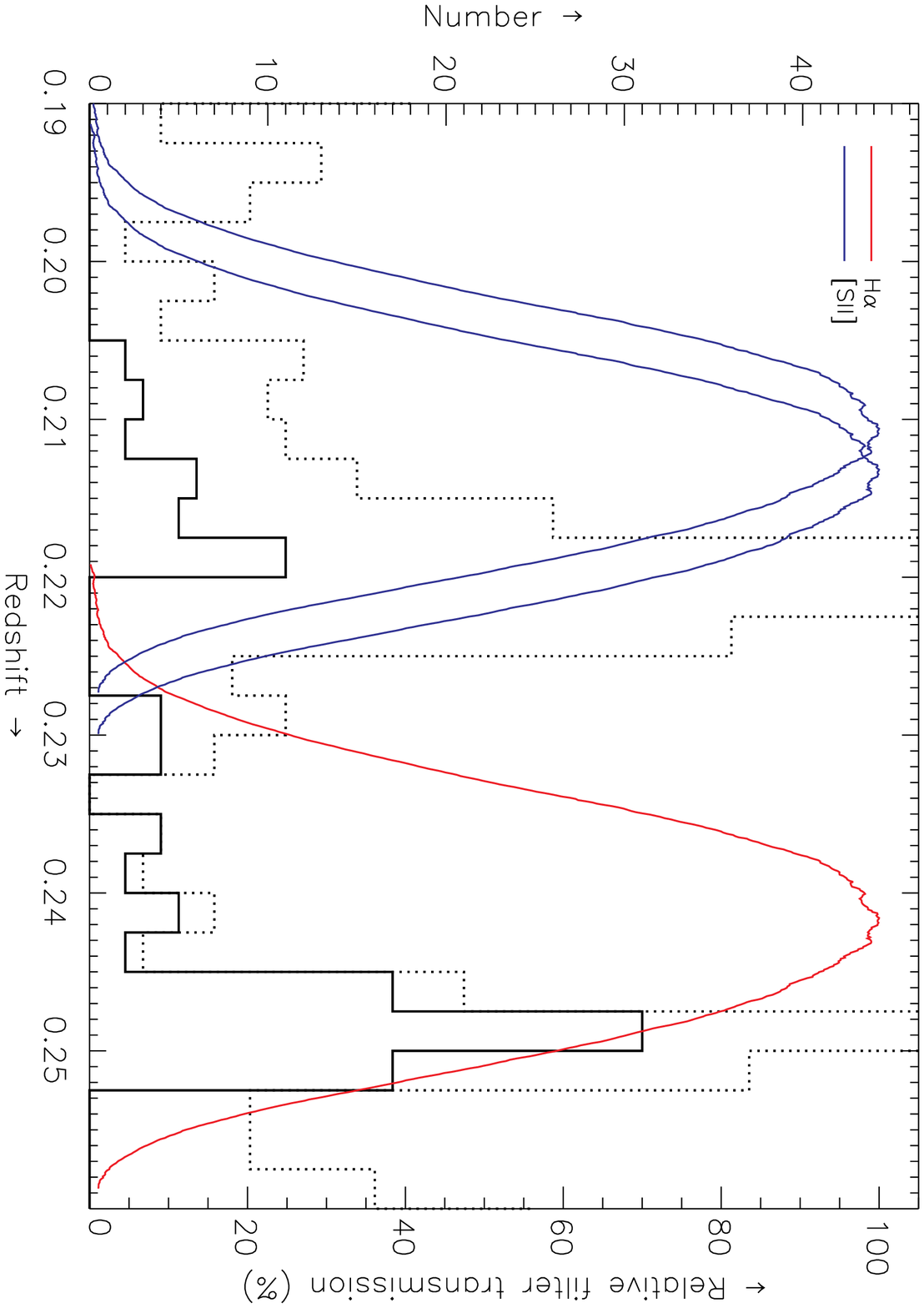}
  \caption{Redshifts from zCOSMOS DR2 of \ha{} candidates of
  \citetalias{Shioya08} ({\it solid histograms}) and all galaxies
  ({\it dotted histogram}), and the relative NB816 filter transmission
  curve for \ha{} ({\it red line}) and both \sii{} lines ({\it blue
  lines}). Note the \citetalias{Shioya08} galaxies within the \sii{}
  sensitive redshift range.}
  \label{fig:zcosmosSII}
\end{figure}

\begin{figure}[tb]
  \centering
  \includeIDLfigPcustom[\columnwidth]{6pt}{16pt}{18pt}{9pt}{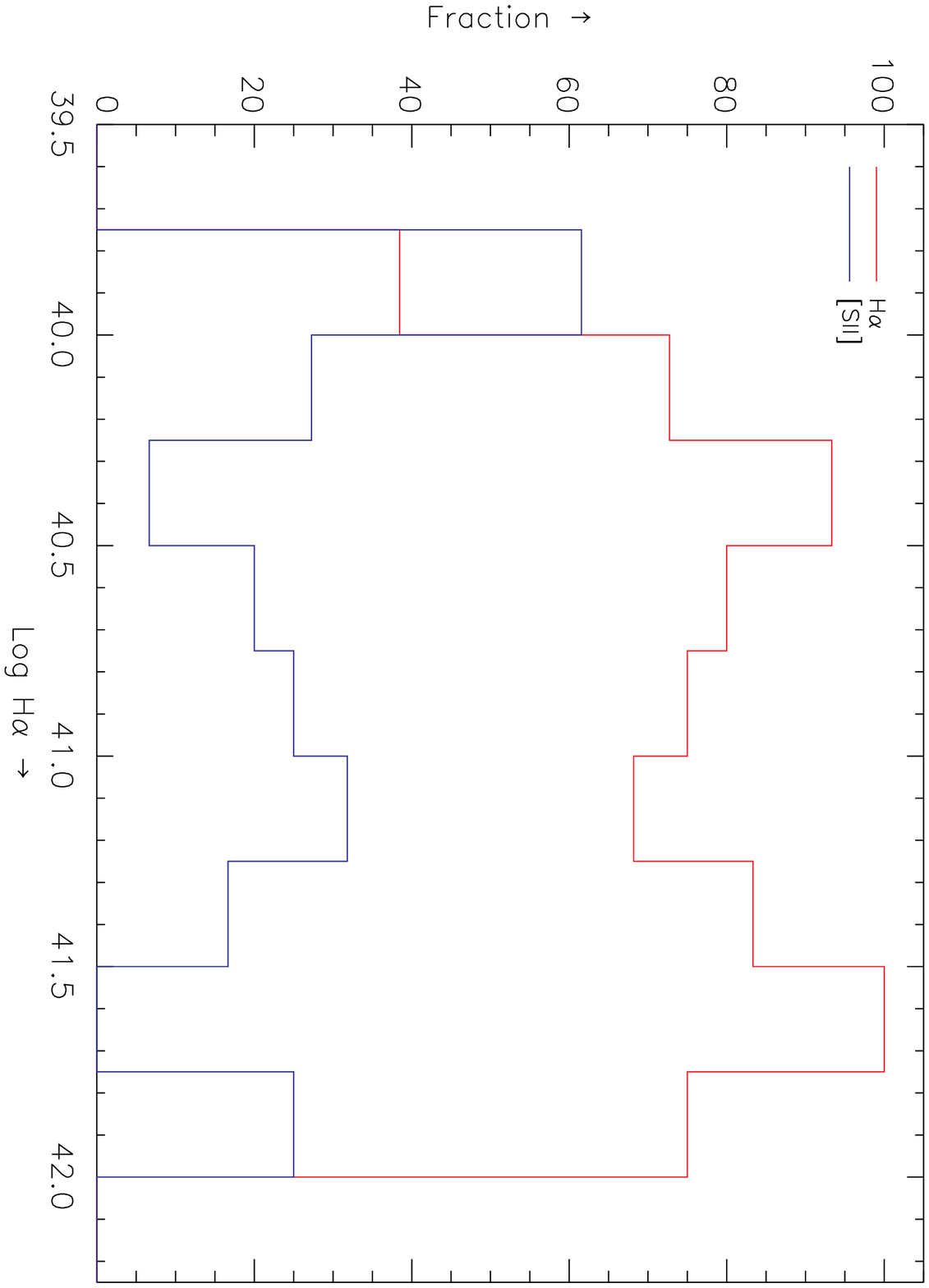}
  \caption{Fraction of candidates with a redshift from zCOSMOS DR2
  corresponding to \ha{} ($z\sim0.24$; {\it red histogram}) or \sii{}
  ($z\sim0.21$; {\it blue histogram}) as a function of \ha{}
  luminosity calculated from the narrowband survey of
  \citetalias{Shioya08}.}
  \label{fig:zcosmosFraction}
\end{figure}

The selection function computed at $z\sim0.24$ using
\citetalias{Shioya08} should lie on top of the SHELS selection
function at $0.200 < z < 0.300$, but it does not. Thus either SHELS
underestimates--or \citetalias{Shioya08} overestimates--the number of
faint \ha{} galaxies. Either the $R$-band magnitude
vs. \ha{}-luminosity relation between SHELS and \citetalias{Shioya08}
must be significantly different, or there is another selection effect
not yet considered. Figure~\ref{fig:shishelscomp} rules out a
different $R$-\ha{} relation. This figure shows that the two surveys
clearly overlap and do not have a significantly different $R$-\ha{}
relation.

A selection effect that removes some galaxies from the SHELS sample is
the OEW$_\mathrm{\ha+\nii} \ge 12$\,\AA{}
criterion. Figure~\ref{fig:combinedLF} shows the effect of this
criterion; it slightly increases the number of galaxies at the
faint-end of the SHELS luminosity function. This bias is, however,
insufficient to explain the differences in the SHELS- and
\citetalias{Shioya08}-based selection functions.

The narrowband survey of \citetalias{Shioya08} may overestimate the
number of fainter \ha{} galaxies. Even though consistent within the
uncertainties, the faint-end slope of the combined luminosity function
at $z \sim 0.24$ is somewhat steeper ($\alpha \sim -1.4$) than at our
lowest redshift bin ($\alpha \sim -1.2$) causing a very steep
selection function at $z \sim 0.24$.

We can determine the redshift of several narrowband-survey candidates
of \citetalias{Shioya08} using zCOSMOS
DR2. Figure~\ref{fig:zcosmosSII} shows the redshifts of the galaxies
from \citetalias{Shioya08} with confirmed redshifts near $z \sim
0.24$. Several galaxies have redshifts outside the wavelength range
where the NB816 filter is sensitive to \ha{} at $z \sim 0.24$ ({\it
red line}). About 25\,\% of the candidates with spectroscopy are at a
lower redshift $z \sim 0.21$. This redshift correspond to the
wavelength range where the NB816 filter is sensitive to the \Sii{}
doublet ({\it blue lines}). These galaxies belong to an overdensity in
the large-scale structure at $z \sim 0.22$ ({\it dotted histogram} in
Figure~\ref{fig:zcosmosSII} and {\it solid histogram} in
Figure~\ref{fig:zcosmos}).

Figure~\ref{fig:zcosmosFraction} shows the fraction of galaxies with
redshifts corresponding to \sii{} or \ha{} as a function of the
\citetalias{Shioya08} \ha{} luminosity. The figure suggests that the
fraction of \sii{} galaxies increases towards fainter
luminosities. This effect could produce an excess of faint \ha{}
galaxies in the \citetalias{Shioya08} survey and thus could explain
the difference in the selection functions implied by the SHELS
simulations and the comparisons of SHELS and \citetalias{Shioya08}.

Color-color selections are sufficient to remove contaminating galaxies
from the narrowband survey at higher redshifts. The contaminants
include \hb{} and \oiii{} at $z \sim 0.6-0.7$, \oii{} at $z \sim 1.2$
and \lya{} at $z \sim 5.7$ for a narrowband survey at $\sim
8150$\,\AA{} \citep[e.g.][]{Fujita03,Ly07}. However, it is impossible
to distinguish \ha{} galaxies from \sii{} galaxies by color
\citep{Westra08}. The contamination is survey dependent because it
depends on the details of the large-scale structure. The
\citetalias{Shioya08} survey is a case where there is a peak in the
redshift distribution exactly where the narrowband survey is sensitive
to \sii{}.

\subsection{Volume dependence}
\label{sec:volumeeffect}
There is a large spread in the parameters determined from different
surveys around $z \sim 0.24$ (Figure~\ref{fig:parevolution})
accompanied by very large uncertainties. All of these surveys
\citep{Fujita03,Hippelein03,Ly07,Westra08} use a single or multiple
narrowband filters over $\sim 300 - 950\,\sq \arcmin$.
\citetalias{Shioya08} uses $5540\,\sq \arcmin$. Typical volumes are
$0.5-1 \times \pow{4}\,\MpcQ$; \citetalias{Shioya08} covers $3 \times
\pow{4}\,\MpcQ$. The smaller volumes are not large enough to constrain
the bright end of the luminosity function. We discussed
\citetalias{Shioya08} in detail in Section~\ref{sec:nbvbb}.

To examine the impact of small volumes, we split SHELS into 16
separate pieces to match the area ($\sim 0.25\,\sq \degr$) of typical
narrowband surveys that probe redshift $\sim
0.24$. Table~\ref{tab:shotnoisetest} gives the median recovered
parameters and the inter-quartile range.

For $\alpha = -1.20$ the recovered parameters are almost identical to
those of the entire field. The inter-quartile range is large, even
when compared to the uncertainties in Table~\ref{tab:schechter}. If we
combine the 16 ``surveys'', we would have to increase the
uncertainties in Table~\ref{tab:schechter} because of the smaller
number of galaxies, i.e. an increase in shot-noise. This uncertainty
easily explains the scatter of the parameters observed at
$z\sim0.24$. It underscores the need for large-volume surveys to
constrain the bright end of the luminosity function.

To constrain $\alpha$ it is more important to have a deep survey and
to span a large range of luminosities rather than to cover a large
area. The data from \citet{Ly07} demonstrate this point. As discussed
in \citetalias{Shioya08}, the data-points from \citetalias{Shioya08}
and \citeauthor{Ly07} are quite similar at the fainter luminosities,
both in slope and in amplitude. Thus, the survey area of
\citeauthor{Ly07}, i.e. $\sim 0.25\,\sq \degr$, can be large enough to
constrain $\alpha$, and $\alpha$ only. Their area (volume) is too
small to determine the bright end of the luminosity function
\citepalias[see also][]{Shioya08} because they do not observe enough
of the rare most-luminous galaxies.

\begin{figure}[tb]
  \centering
  \includeIDLfigPcustom{14pt}{16pt}{28pt}{5pt}{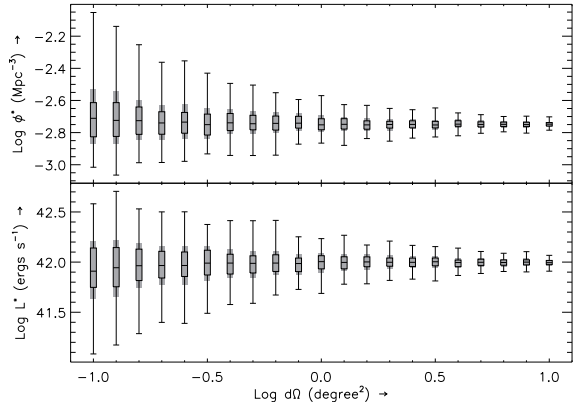}
  \caption{Box and whisker plot for the simulated surveys as a
  function of area. The surveys cover $0.233 < z < 0.251$ and have a
  limiting flux of $\pow{-15.5}$\,\fluxunits{}. The gray box indicates
  the 1\,$\sigma$ around the mean, the dash indicates the median, the
  boxes indicates the inter-quartile range, and the whiskers indicate
  minimum and maximum values of recovered Schechter parameters (using
  the STY-method) from the simulations.}
  \label{fig:volumetest}
\end{figure}

To estimate the area required to constrain $L^*$ and $\phi^*$, we
simulate many observed galaxies given a specific luminosity function
at $0.233 < z < 0.251$ for different sized areas. We fit the
parameters with fixed $\alpha = -1.20$. {\it On average}, the
parameters are very well recovered. However, for the smaller areas the
spread in the recovered parameters is large. We show the 1\,$\sigma$
spread around the mean, the median, the inter-quartile range, and
minimum and maximum values of the recovered values for each area in
Figure~\ref{fig:volumetest}.

\begin{deluxetable}{lcccccc}
  \tablewidth{0pt}

  \tablecaption{Median, upper and lower quartile range of Schechter
  parameters for 0.25\,\sq{}\degr{} subsets using
  SHELS.\label{tab:shotnoisetest}}

  \tablehead{
    \colhead{} & \multicolumn{3}{c}{fixed $\alpha$} & \multicolumn{3}{c}{unconstrained $\alpha$}\\
    \colhead{redshift range} & \colhead{$\alpha$} & \colhead{$\log L^*$} & \colhead{$\log \phi^*$} & \colhead{$\alpha$} & \colhead{$\log L^*$} & \colhead{$\log \phi^*$}
  }

  \startdata

$0.010 < z < 0.100$ & $-1.20$ & $41.60 _{-0.17} ^{+0.23}$ & $-2.79 _{-0.13} ^{+0.06}$ & $-1.18 _{-0.10} ^{+0.19}$ & $41.58 _{-0.35} ^{+0.23}$ & $-2.76 _{-0.34} ^{+0.26}$\\
$0.100 < z < 0.200$ & $-1.20$ & $42.06 _{-0.09} ^{+0.10}$ & $-2.96 _{-0.13} ^{+0.07}$ & $-0.62 _{-0.32} ^{+0.10}$ & $41.63 _{-0.22} ^{+0.18}$ & $-2.57 _{-0.05} ^{+0.28}$\\
$0.200 < z < 0.300$ & $-1.20$ & $42.53 _{-0.19} ^{+0.13}$ & $-3.28 _{-0.12} ^{+0.08}$ & $-0.61 _{-0.16} ^{+0.15}$ & $42.07 _{-0.21} ^{+0.21}$ & $-2.75 _{-0.07} ^{+0.07}$\\
$0.300 < z < 0.377$ & $-1.20$ & $42.82 _{-0.05} ^{+0.04}$ & $-3.56 _{-0.09} ^{+0.05}$ & $-0.48 _{-0.12} ^{+0.17}$ & $42.27 _{-0.15} ^{+0.04}$ & $-2.96 _{-0.24} ^{+0.02}$\\
$0.233 < z < 0.251$ & $-1.20$ & $42.40 _{-0.22} ^{+0.22}$ & $-3.28 _{-0.09} ^{+0.07}$ & $-0.22 _{-0.44} ^{+0.21}$ & $41.91 _{-0.27} ^{+0.11}$ & $-2.79 _{-0.63} ^{+0.20}$\\

  \enddata

\end{deluxetable}

If we assume that 10\,\% is an acceptable uncertainty for a parameter
($\sim 0.04$ in dex), then the survey area required is $\sim 3\,\sq
\degr$. Surveys like \citet{Fujita03}, \citet{Ly07} and
\citet{Westra08} at $z \sim 0.24$ are thus not large enough to
constrain the bright end of the luminosity
function. \citetalias{Shioya08} is a factor of two shy of this area;
SHELS is larger.

Hence, combining the \citetalias{Shioya08} and SHELS data
(Section~\ref{sec:combinedlf}) is an excellent way to constrain the
faint and bright end of the luminosity function simultaneously.

\begin{figure*}[tb]
  \centering
  \includeIDLfigPcustom[0.75\textwidth]{10pt}{16pt}{28pt}{10pt}{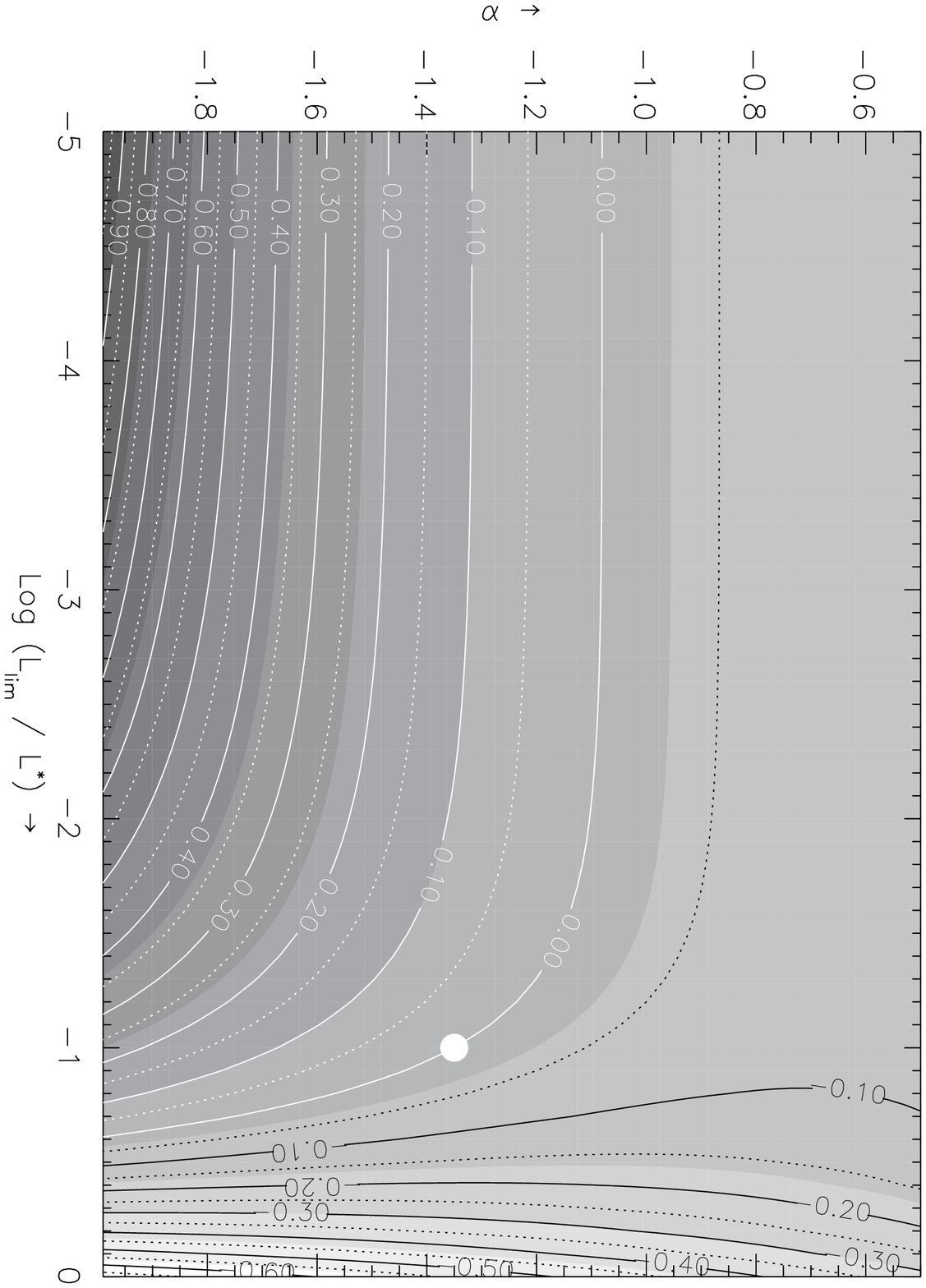}
  \caption{Logarithm of the total luminosity density evaluated at
    ($\log \frac{L_\mathrm{lim}}{L^*}$, $\alpha$) divided by the
    luminosity density at ($\log \frac{L_\mathrm{lim}}{L^*}$,
    $\alpha$) = (-1, -1.35) ({\it white filled circle}). The black
    lines indicate negative values.}
  \label{fig:lumcuttest}
\end{figure*}

\section{Star formation density}
\label{sec:sfd}
We determine the star formation density ($\dot{\rho}$ in \sfDens{})
from the integrated \ha{} luminosity density for each
redshift range. We use the conversion from \ha{} luminosity to star
formation rate from \citet{Kennicutt98} for Case B recombination and
$T_e = \pow{4}$\,K
\begin{equation}
\mathrm{SFR} = 7.9 \times \pow{-42} L_\mathrm{\ha{}},
\end{equation}
where SFR in \Msunyr{} and $L_\mathrm{\ha{}}$ in \ergs{}. We determine
the \ha{} luminosity density for $L \ge L_\mathrm{lim}$ from the
parameters of the Schechter function using
\begin{equation}
  \mathcal{L} = \phi^* L^* \Gamma(\alpha+2,
  \frac{L_\mathrm{lim}}{L^*}),
  \label{eq:lumdens}
\end{equation}
where $\Gamma$ is the incomplete gamma function. 

The choice $L_\mathrm{lim} = 0$ affects the integrated luminosity
density\footnote{$L_\mathrm{lim} = 0$ reduces Eq~\eqref{eq:lumdens} to
$\mathcal{L} = \phi^* L^* \Gamma(\alpha+2)$, where $\Gamma$ is the
complete gamma function.}. Figure~\ref{fig:lumcuttest} shows the
logarithm of the total luminosity density evaluated at ($\log
\frac{L_\mathrm{lim}}{L^*}$, $\alpha$) divided by the luminosity
density at ($\log \frac{L_\mathrm{lim}}{L^*}$, $\alpha$) = (-1,
-1.35). We can thus determine the effect of using different limiting
luminosities on the total luminosity density. For example, using
$L_\mathrm{lim} = 0$ for $\alpha = -1.35$ rather than $L_\mathrm{lim}
= 0.1 L^*$ gives a difference of $\pow{0.12 - 0.00} = 1.31$, i.e. an
increase of 30\,\%. These effects are obviously more severe for
steeper values of $\alpha$. When comparing surveys of different
depths, one needs to be careful about extrapolations of the \ha{}
luminosity function (Schechter function) to very low star formation
rates, especially for steep $\alpha$.

\begin{deluxetable}{lcccccc}
  \tablewidth{0pt}

  \tablecaption{Star formation density.\label{tab:sfd}}

  \tablehead{
    \colhead{} & \colhead{} & \multicolumn{2}{c}{$\log \dot{\rho}$} & \multicolumn{2}{c}{$\log \dot{\rho}$ with $\log L_\mathrm{lim} = 40$}\\
    \colhead{redshift range} & \colhead{$\log L_\mathrm{lim}$\tablenotemark{a}} & fixed $\alpha$ & unconstrained $\alpha$ & fixed $\alpha$ & unconstrained $\alpha$
  }

  \tablewidth{0pt}

  \startdata

  $0.010 < z < 0.100$\tablenotemark{b} & 37.84 & $-2.18 \pm 0.10$ & $-2.19 \pm 0.17$ & $-2.20 \pm 0.10$ & $-2.21 \pm 0.17$\\
  $0.100 < z < 0.200$ & 39.89 & $-1.92 \pm 0.09$ & $-1.92 \pm 0.12$ & $-1.93 \pm 0.09$ & $-1.92 \pm 0.12$\\
  $0.200 < z < 0.300$ & 40.55 & $-1.82 \pm 0.05$ & $-1.81 \pm 0.10$ & $-1.81 \pm 0.05$ & $-1.81 \pm 0.10$\\
  $0.300 < z < 0.377$ & 40.95 & $-1.81 \pm 0.03$ & $-1.82 \pm 0.08$ & $-1.80 \pm 0.03$ & $-1.81 \pm 0.08$\\
  $0.233 < z < 0.251$\tablenotemark{c} & \multicolumn{3}{c}{} & \nodata & $-1.86 \pm 0.13$\\

  \enddata

  \tablenotetext{a}{$L_\mathrm{lim} = 4 \pi D_L ^2 (z_\mathrm{low})$}

  \tablenotetext{b}{The redshift range $0.010 < z < 0.100$ covers an
  atypical under-dense region (Section~\ref{sec:fieldselection}).}

  \tablenotetext{c}{Combined SHELS and \citetalias{Shioya08} result.}

  \tablecomments{We use the Schechter parameters determined for the
  pure star forming galaxies in Table~\ref{tab:schechter}. We
  calculate the uncertainties using standard uncertainty propagation
  for Eq.~(\ref{eq:lumdens}) and the uncertainties in
  Table~\ref{tab:schechter}. $\dot{\rho}$ is in \sfDens{}.}
\end{deluxetable}

\begin{figure*}[tb]
  \centering
  \includeIDLfigPcustom[0.75\textwidth]{14pt}{16pt}{10pt}{5pt}{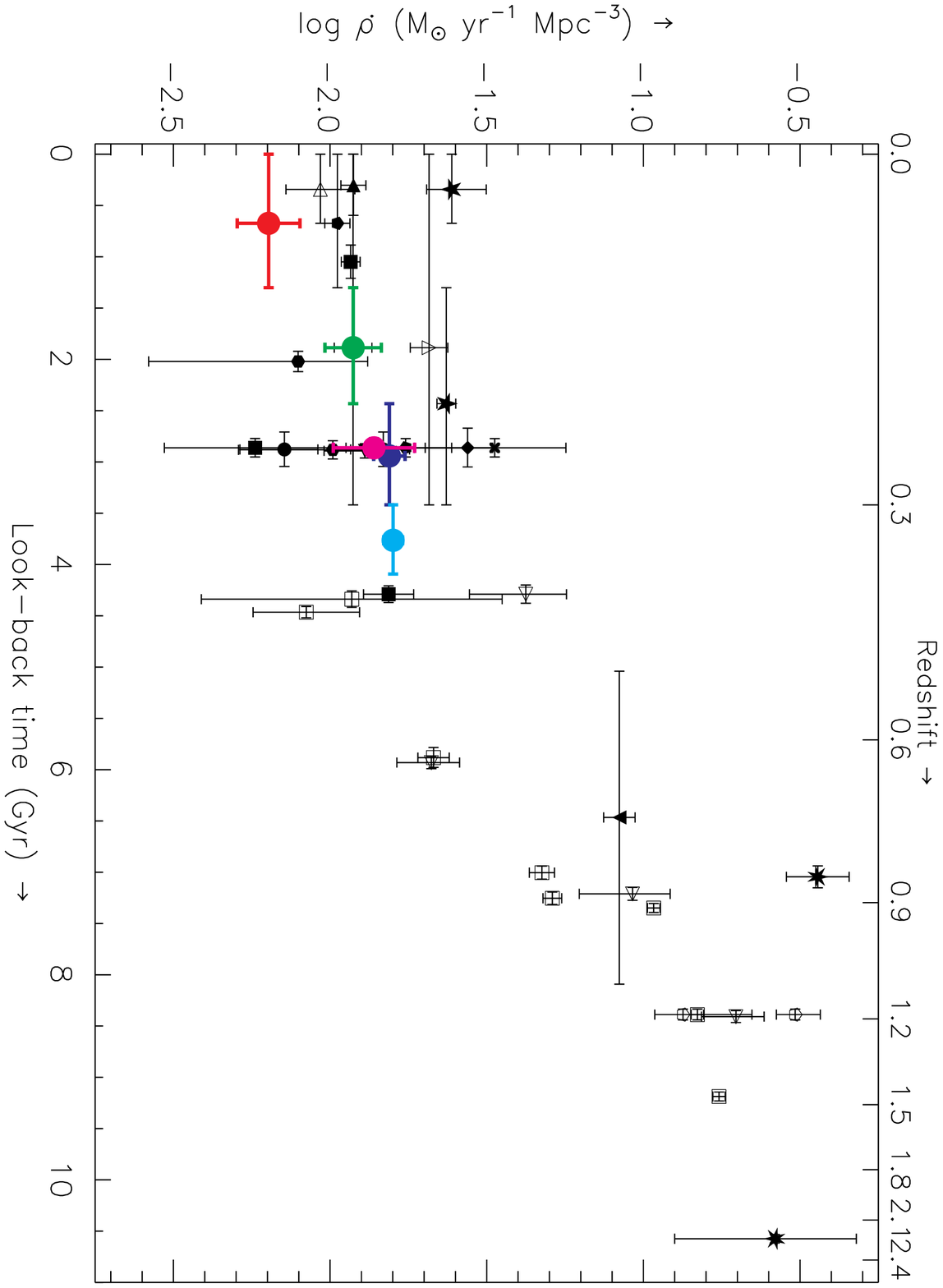}
  \includeIDLfigPcustom[0.24\textwidth]{85pt}{435pt}{40pt}{70pt}{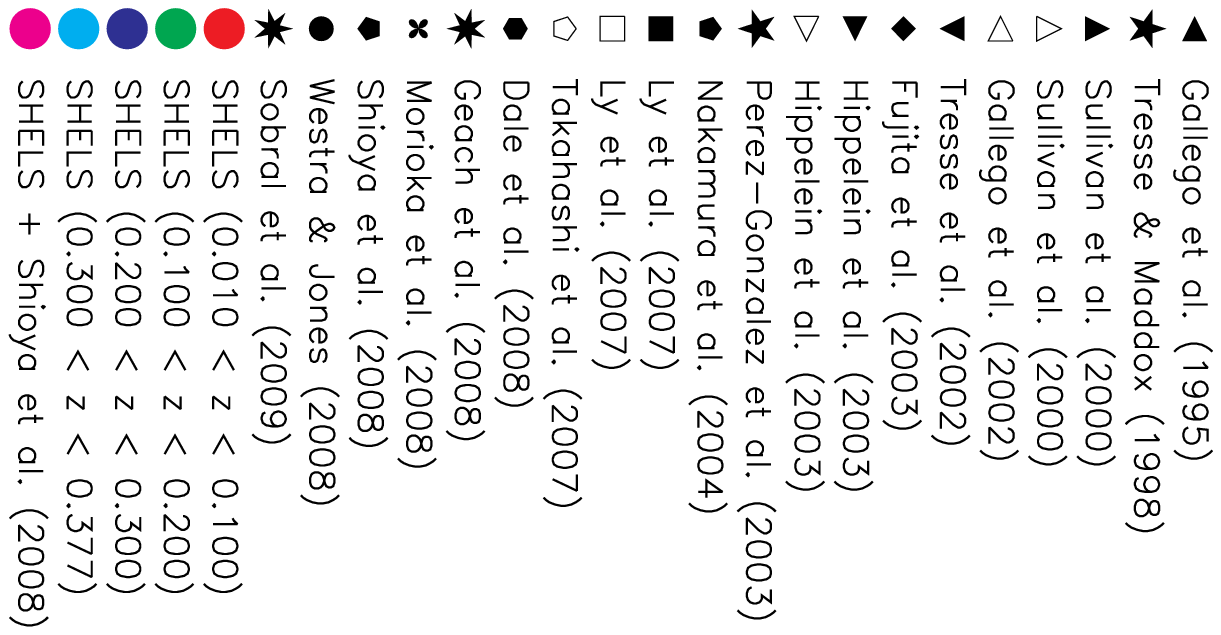}
  \caption{Star formation density as a function of look-back time and
  redshift for SHELS ({\it red, green, blue, and cyan large solid
  circles}) compared with other surveys using the \ha{} line ({\it
  solid symbols}), or either \oii{} or \oiii{} lines ({\it open
  symbols}) as star formation indicator. We also indicate the combined
  SHELS and \citetalias{Shioya08} point ({\it solid large magenta
  circle}). We calculate the star formation density using the
  Schechter parameters of each survey to a limiting star formation
  rate of 0.079\,\Msunyr{} (corresponds to $L_\ha{} =
  \pow{40}$\,\ergs{}) to reduce the systematic uncertainty from
  extrapolation to $L = 0$\,\ergs{}.}
  \label{fig:sfd}
\end{figure*}

Table~\ref{tab:sfd} lists the star formation densities and
uncertainties for SHELS down to the luminosity limit of the
appropriate redshift bin. We also show the star formation densities
down to $\log L_\mathrm{lim} = 40.00$ corresponding to a star
formation rate of 0.079\,\Msunyr{} for comparison with other surveys
(Figure~\ref{fig:sfd}). We choose this value for all surveys because
most surveys either reach this star formation rate, or the required
extrapolation is modest. The solid symbols in Figure~\ref{fig:sfd}
represent surveys with star formation densities derived from the \ha{}
line; the open symbols come from either the \oii{} or \oiii{}
line. Figure~\ref{fig:sfd} shows that the star formation density for
other surveys at $0.200 < z < 0.300$ is consistent with the star
formation density determined from the combined luminosity function of
SHELS and \citetalias{Shioya08}.

Figure~\ref{fig:sfd} also shows a clear increase in the star formation
density with increasing redshift. However, our lowest redshift point
($0.010 \le z < 0.100$) lies below surveys at similar redshifts. This
underestimate occurs because our field was selected against low
redshift clusters. This survey is thus an underdense region at low
redshifts and the star formation density is probably correspondingly
underestimated.

Because we use the integrated Schechter function to determine the star
formation density, the arguments in Section~\ref{sec:volumeeffect} for
the Schechter parameters $L^*$ and $\phi^*$, are valid for the star
formation density. The median, upper and lower quartile range for the
star formation density for 0.25\,\sq{}\degr{} subsets are in
Table~\ref{tab:sfdnoise}. Again, the recovered star formation density
of a 0.25\,\sq{}\degr{} subset is almost identical to that of the
entire field, but the standard deviation in the star formation density
is very large (almost a factor of 2). The large uncertainty mainly
results from the scatter in $L^*$ and $\phi^*$. Again, combined with
the increased uncertainties resulting from increased shot-noise, the
spread in star formation densities at narrow redshift slices can
easily be explained by sampling a volume that is too small.

\begin{deluxetable}{lcc}
  \tablewidth{0pt}

  \tablecaption{Median, upper and lower quartile range of the star
  formation density for 0.25\,\sq{}\degr{} subsets using
  SHELS.\label{tab:sfdnoise}}

  \tablehead{
    \colhead{} & \multicolumn{2}{c}{$\log \dot{\rho}$}\\
    \colhead{redshift range} & fixed $\alpha$ & unconstrained $\alpha$ 
  }

  \startdata
$0.010 < z < 0.100$ & $-2.25 _{-0.23} ^{+0.12}$ & $-2.30 _{-0.22} ^{+0.12}$\\
$0.100 < z < 0.200$ & $-1.96 _{-0.15} ^{+0.13}$ & $-1.97 _{-0.15} ^{+0.25}$\\
$0.200 < z < 0.300$ & $-1.84 _{-0.13} ^{+0.16}$ & $-1.85 _{-0.13} ^{+0.16}$\\
$0.300 < z < 0.377$ & $-1.78 _{-0.05} ^{+0.03}$ & $-1.82 _{-0.19} ^{+0.03}$\\
$0.233 < z < 0.251$ & $-1.95 _{-0.26} ^{+0.23}$ & $-1.96 _{-0.26} ^{+0.23}$\\

  \enddata

  \tablecomments{The values for the star formation density are
  integrated down to $\log L_\mathrm{lim} = 40$.}

\end{deluxetable}

\section{Physical properties of star forming galaxies}
\label{sec:properties}

\subsection{Stellar population age}

\begin{figure}[tb]
  \centering
  \includeIDLfigPcustom{15pt}{16pt}{28pt}{20pt}{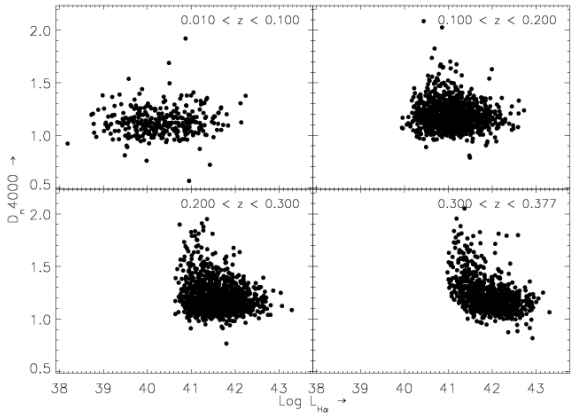}
  \caption{$D_n{4000}$ as a function of \ha{} luminosity for each of
    the four redshift bins for pure star forming \ha{} emitting galaxies.}
  \label{fig:d4000}
\end{figure}

In star forming galaxies the \ha{} emission originates from gas
surrounding the young stars. The spectrum from an actively star
forming galaxy is dominated by the light emitted by these young
stars. Figure~\ref{fig:d4000} shows $D_n{4000}$, the ratio of the
continuum red- and bluewards of the $H+K$ break and an indicator of
the age of the stellar population \citep{Balogh99,Bruzual83}, as a
function of the \ha{} luminosity for pure star forming galaxies. A low
$D_n{4000}$ \citetext{$D_n{4000} \lesssim 1.44$; D.~F.~Woods et
al. 2010; in preparation} indicates a young stellar population. The
majority of the \ha{} emitting galaxies contain a young stellar
population.

\subsection{Galaxy-galaxy interaction}

\citet{Sobral09} find that the fraction of mergers rises with
increasing luminosity particularly around $L^*$. Some of the SHELS
galaxies are quite luminous in \ha{} indicating they are undergoing a
starburst. \citet{Barton00} find that a close pass of two galaxies can
initiate a starburst. Following \citeauthor{Sobral09}, we examine the
SHELS data to look for evidence of the impact of interactions on the
\ha{} luminosity function. Thus, we focus on galaxies that may have
(or may have had) a recent encounter with another galaxy. We determine
whether each galaxy has an apparently nearby ``neighbor''.

\begin{figure*}[tb]
  \centering
  \includeIDLfigPcustom[0.75\textwidth]{8pt}{16pt}{28pt}{20pt}{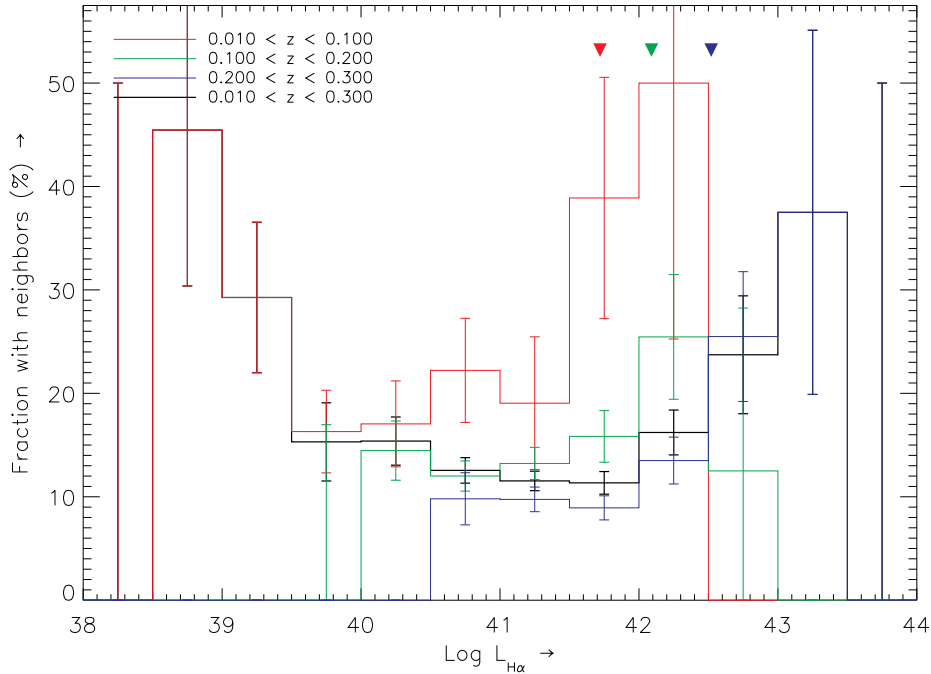}
  \caption{Fraction of pure star forming galaxies with one or more
  neighbors for each redshift bin ({\it colored histograms}) as a
  function of redshift and the fraction for $0.01 < z < 0.30$ ({\it
  black thick histogram}). Colored triangles indicate $L^*$
  (determined with $\alpha = -1.20$) for each redshift range.}
  \label{fig:friendshisto}
\end{figure*}

A galaxy has a neighbor when the velocity difference (corrected for
redshift) between the two galaxies is $\le 500\,\kms$, and their
projected separation is $\le 100$\,kpc. These values are a standard
definition of galaxy pairs
\citep[e.g.][]{Barton00,Patton00,Lin04,Woods07,Park09}. We include the
somewhat deeper SHELS catalog to look for neighboring galaxies (see
Section~\ref{sec:fieldselection}). This catalog contains spectra of
galaxies with magnitudes $20.3 < R < 20.6$ where the spectroscopy is
52\,\% complete. The fraction of all our pure star forming galaxies
that have a neighbor is 15.3\,\% (547 out of 3565)\footnote{If we
decrease our projected separation criterion to 50\,kpc, the fraction
drops with a factor of $\sim$2 to 7.4\,\% (265 out of 3565). The
fractions in Figure~\ref{fig:friendshisto} scale with roughly the same
factor within the uncertainties. Our conclusions are not affected by
the choice of projected separation.}.

Figure~\ref{fig:friendshisto} shows the fraction of galaxies with one
or more neighbors as a function of \ha{} luminosity for the lowest
three redshift bins ({\it colored histograms}) and for the three
redshift ranges combined ({\it thick black histogram}). The fraction
is always a lower limit; deeper spectroscopy might reveal only more
neighbors of a galaxy, never fewer.

It is striking that the fraction of galaxies with neighbors increases
around $L^*$ (and for the lowest redshift bin towards the lowest \ha{}
luminosities). This result agrees with a rise in the fraction of
mergers with increasing \ha{} luminosity found by
\citet{Sobral09}. The interesting question for galaxy evolution is
whether the location of the increase determines $L^*$, or whether
$L^*$ determines the location of the increase.

\begin{figure*}[tb]
  \centering
  \includeIDLfigPcustom[0.75\textwidth]{8pt}{16pt}{28pt}{20pt}{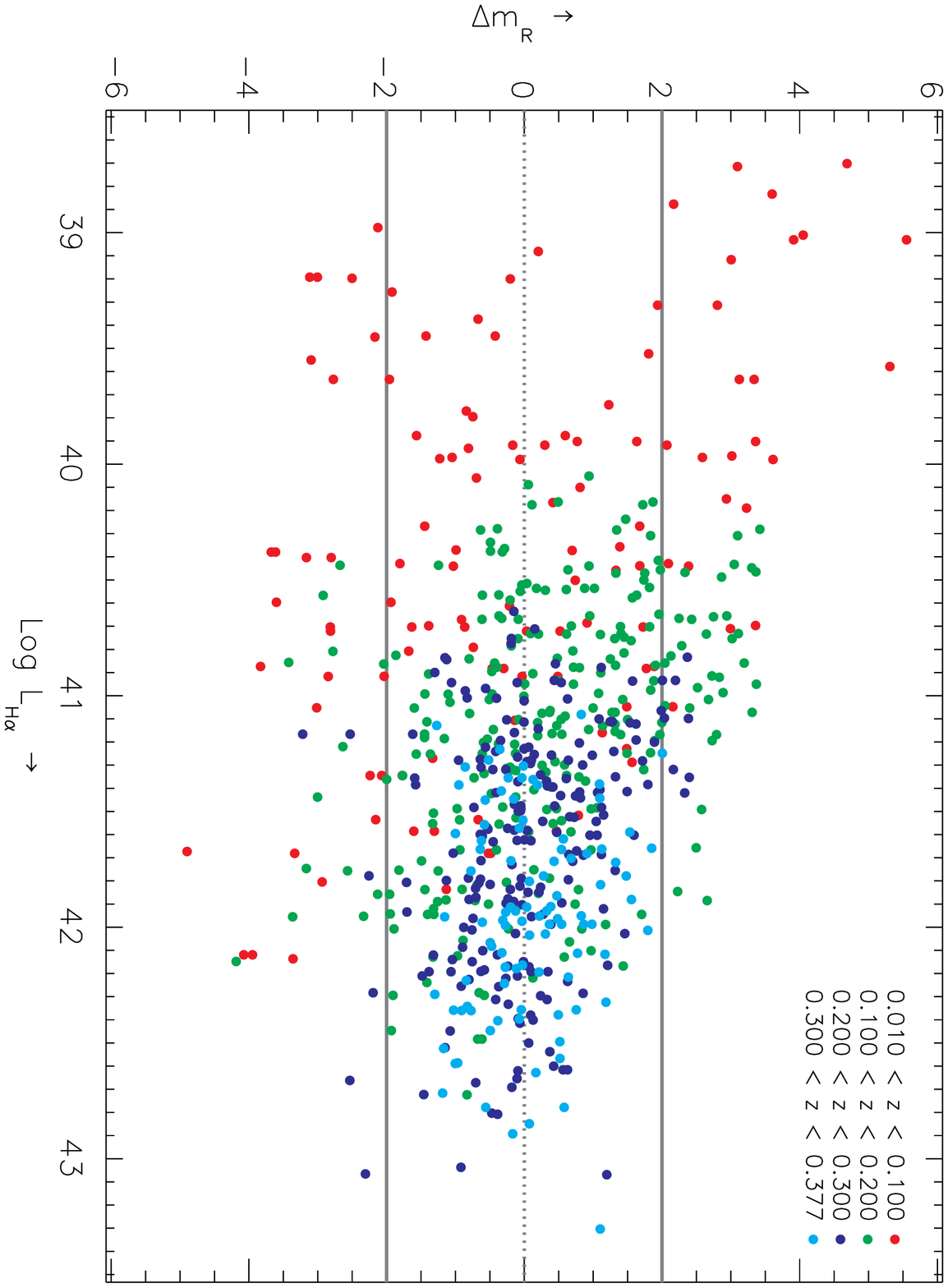}
  \caption{Magnitude difference between the galaxy and its neighbor as
  a function of \ha{} luminosity for pure star forming \ha{} emitting
  galaxies. The solid lines show $|\Delta m_R| = 2$, the demarcation
  between major and minor interactions. Galaxies above the dotted line
  are fainter than their neighbor, galaxies below are more luminous.}
  \label{fig:minormajor}
\end{figure*}

To investigate this behavior further, we investigate the magnitude
difference between the galaxy and its
neighbor(s). Figure~\ref{fig:minormajor} shows the \ha{} luminosity as
a function of the magnitude difference $\Delta m_R = R_\mathrm{galaxy}
- R_\mathrm{neighbor}$\footnote{For neighbors in the $R_\mathrm{tot} >
20.3$ catalog we assumed $R = 20.3$. Thus, $\Delta m_R =
R_\mathrm{galaxy} - 20.3$ for these galaxies.}. We also indicate the
demarcation between minor and major pairs, i.e. $|\Delta m| = 2$
\citep[e.g.][]{Woods07}.

Luminous \ha{} galaxies with neighbors tend to be mostly part of a
major pair, and to a lesser extent the more luminous galaxy of a minor
pair; faint \ha{} galaxies with neighbors can be part of a major or
minor pair. However, when faint \ha{} galaxies are part of a minor
pair, they tend to be the fainter (smaller) galaxy. The behavior in
Figure~\ref{fig:minormajor} is consistent with the picture of
interaction-induced star formation. The increase in the fraction
around $L^*$ implies that galaxy-galaxy interactions are important for
the increase of the \ha{} luminosity in these galaxies.

\section{Summary and conclusion}
\label{sec:summary}

We use the Smithsonian Hectospec Lensing Survey (SHELS) to study \ha{}
emitting galaxies. SHELS is complete to $R_\mathrm{tot} = 20.3$ over a
large 4\,\sq{}\degr{} area. This area yields a large enough volume to
study the bright end of the \ha{} luminosity function as a function of
redshift.

We determine the \ha{} flux and attenuation from the SHELS
spectroscopy. We also identify galaxies that host AGNs or are
composites.

We combine the strengths of two surveys, the breadth of SHELS (to
constrain the bright-end of the luminosity function) and the depth of
the narrowband survey of \citetalias{Shioya08} (to determine the faint
end slope of the luminosity function), to determine a well-constrained
\ha{} luminosity function at $z \sim 0.24$. A narrowband survey goes
deep over a limited field of view to cover the faint end of the
luminosity function. A broadband selected spectroscopic survey can
easily cover a larger volume to probe the bright end of the luminosity
function. The resulting Schechter parameters are consistent with
\citetalias{Shioya08} within their uncertainties.

We determine the \ha{} luminosity function from SHELS for four
redshift intervals over $0.010 < z < 0.377$. The lowest redshift
interval ($0.010 < z < 0.100$) covers an atypical underdense region
due to field selection. The characteristic luminosity $L^*$ increases
as a function of redshift ($\Delta \log L^* = 0.84$ over $0.100 < z <
0.377$).

The star formation density also increases with increasing redshift
($\Delta \log \dot{\rho} = 0.11$ over $0.010 < z < 0.377$). The star
formation rate from the combined luminosity function of SHELS and
\citetalias{Shioya08} is consistent with that of SHELS alone at $0.200
< z < 0.300$.

The fraction of galaxies with neighbors increases by a factor of $2-5$
around $L^*$ for the most luminous star forming galaxies at each
redshift, similar to \citet{Sobral09}. The fraction appears to also
increase towards fainter \ha{} luminosity as a result of interactions
in minor pairs. We conclude that triggered star formation is important
for both the highest and lowest luminosity \ha{} galaxies.

The future of surveys for star forming galaxies is a combination of a
large-area spectroscopic survey combined with very deep narrowband
imaging. However, the narrowband imaging requires extensive test
spectroscopy because the impact of large-scale structure with respect
to the filter response is unknown a priori. The combination of methods
can constrain and remove the scatter in the star formation density as
a function of redshift. The combination also allows a secure
determination of the shape of the luminosity function over a large
luminosity range.

\section*{Acknowledgments}

We thank Christy Tremonti for providing her continuum subtraction
routine, Anil Seth for suggesting the usage of the routine to
compensate for the underlying stellar absorption, Antonaldo Diaferio
for discussions on Schechter function fitting, Warren Brown, Scott
Kenyon, and Deborah Woods for useful discussions. We are grateful for
the contributions of the members of the MMT Observatory and the
Telescope Data Center of the CfA. EW acknowledges the Smithsonian
Institution for the support of his post-doctoral fellowship.

We appreciate the thorough reading of this manuscript by an anonymous
referee whose report has helped to improve the paper.

Observations reported here were obtained at the MMT Observatory, a
joint facility of the Smithsonian Institution and the University of
Arizona. The SDSS is managed by the Astrophysical Research Consortium
for the Participating Institutions. The zCOSMOS observations were made
with ESO Telescopes at the La Silla or Paranal Observatories under
programme ID 175.A-0839.

\bibliographystyle{apj}
\begin{footnotesize}
\bibliography{ms.bib}
\end{footnotesize}

{\it Facilities:} \facility{MMT (Hectospec)}

\end{document}